\documentclass[superscriptaddress,11pt,nofootinbib]{revtex4-1}

\usepackage{bbm,bm,graphicx,mathtools,color,hyperref,slashed}
\usepackage{amssymb,amsmath}
\usepackage[T1]{fontenc}
\usepackage{enumerate}
\usepackage[all]{xy}

\newcommand{\id}{{\bf 1}}

\newcommand{\be}{\begin{equation}}
\newcommand{\ee}{\end{equation}}
\newcommand{\ba}{\begin{align}}
\newcommand{\ea}{\end{align}}
\newcommand{\bea}{\begin{eqnarray}}      
\newcommand{\eea}{\end{eqnarray}}

\newcommand{\ve}{\varepsilon}

\newcommand{\im}{\mathrm{i}}

\newcommand{\rme}{\mathrm{e}}

\newcommand{\PpT}{(P^+_\mu)^\mathrm{T}}
\newcommand{\PmT}{(P^-_\mu)^\mathrm{T}}

\newcommand{\T}{\mathrm{T}}

\newcommand{\A}{X}
\newcommand{\B}{Y}

\begin{document}

\title{
Varieties and properties of central-branch Wilson fermions 
}

\author{Tatsuhiro Misumi}
\email{misumi(at)phys.akita-u.ac.jp}
\address{Department of Mathematical Science, Akita University,  Akita 010-8502, Japan}
\address{Research and Education Center for Natural Sciences, Keio University, 4-1-1 Hiyoshi, Yokohama, Kanagawa 223-8521, Japan}

\author{Jun Yumoto}
\email{m8019309(at)s.akita-u.ac.jp}
\affiliation{Department of Mathematical Science, Akita University, Akita 010-8502, Japan}

\begin{abstract}
We focus on the four-dimensional central-branch Wilson fermion, which makes good use of six species at the central branch of the Wilson Dirac spectrum and possesses the extra $U(1)_{\overline V}$ symmetry. 
With introducing new insights we discuss the prohibition of additive mass renormalization for all the six species, SSB of $U(1)_{\overline V}$ in strong-coupling QCD, the absence of the sign problem, and the usefulness for many-flavor QCD simulation.
We then construct several varieties of the central-branch fermions and study their properties. In particular, we investigate the two-flavor version, where the Dirac spectrum has seven branches and two species live at the central branch. Although the hypercubic symmetry is broken, the other symmetries are the same as those of the original one. We study this setup in terms of lattice perturbation theory, strong-coupling QCD, the absence of sign problem, and the parameter tuning for Lorentz symmetry restoration.
By comparing the properties of the original and two-flavor version, we find that the existence of hypercubic symmetry as well as $U(1)_{\overline V}$ is essential for the absence of additive mass renormalization of all the central-branch species.
As the other two-flavor version, we investigate the central-branch staggered-Wilson fermion, which is obtained from the eight-flavor central-branch Wilson fermion via spin diagonalization.
We argue that it is free from any additive mass renormalization and is regarded as a minimally doubled fermion with less symmetry breaking. 
\end{abstract}

\maketitle

\newpage

\tableofcontents

\newpage


\section{Introduction}
\label{sec:Intro}

In the last four decades theoretical physicists have successfully been studying non-perturbative aspects of quantum field theories including Yang-Mills theory and Quantum Chromodynamics (QCD) by use of lattice gauge theory \cite{Wilson:1974sk, Creutz:1980zw}.  
Although the accomplishments obtained by the technique spread through a broad field of particle and nuclear physics, the research of lattice fermion formulations still has lots of topics left to be investigated \cite{Karsten:1980wd, Nielsen:1980rz, Nielsen:1981xu, Nielsen:1981hk}.
There are several unsolved problems, including the reconcilement of a desirable number of flavors, chiral symmetry and numerical efficiency, the realization of a single Weyl fermion, the sign problem of the quark determinant, etc. These problems motivate us to continue to study lattice fermions from various viewpoints.

Although practically useful fermion formulations have been developed so far, all of them have their individual shortcomings. The Wilson fermion\footnote{
The Wilson fermion with negative mass $-2r < m < 0$ corresponds to a nontrivial symmetry-protected-topological (SPT) phase, where the transition to another SPT phase requires the gap to be closed. It means that a massless Domain-wall fermion appears at the boundary.
}
 with explicit chiral symmetry breaking results in an additive mass renormalization and $O(a)$ discretization errors \cite{Wilson:1975id}. Domain-wall or overlap fermions, which produce a single-flavor chiral-symmetric fermion with errors starting from $O(a^{2})$, lead to rather expensive simulation algorithms \cite{Kaplan:1992bt, Shamir:1993zy, Furman:1994ky, Neuberger:1998wv,Ginsparg:1981bj}. Staggered fermion keeps a $U(1)$ subgroup of chiral symmetry and eliminate $O(a)$ errors, where the degeneracy of four flavors requires the rooting trick for realistic $(2+1)$-flavor QCD \cite{Kogut:1974ag, Susskind:1976jm, Kawamoto:1981hw,Sharatchandra:1981si,Golterman:1984cy,Golterman:1985dz,Kilcup:1986dg}.

In these years, several new approaches to lattice fermion formulations have been investigated:
The Wilson term in Wilson fermion can be generalized to ``flavored-mass terms". 
Based on this generalization of the Wilson term one can construct various cousins of Wilson fermions. 
Similar flavored-mass terms can be introduced into staggered fermions \cite{Golterman:1984cy, Adams:2009eb,Adams:2010gx,Hoelbling:2010jw} and one obtain, what is called, staggered-Wilson fermions \cite{Adams:2009eb,Adams:2010gx,Hoelbling:2010jw,deForcrand:2011ak,Creutz:2011cd,Misumi:2011su,Follana:2011kh,deForcrand:2012bm,Misumi:2012sp,Misumi:2012eh,Durr:2013gp,Hoelbling:2016qfv,Zielinski:2017pko}.
It can be applied to lattice simulations as another version of Wilson fermion or an overlap kernel.
Simple generalizations of Wilson fermion based on flavored-mass terms and their application as the overlap kernel are also intensively investigated \cite{Bietenholz:1999km,Creutz:2010bm,Durr:2010ch,Durr:2012dw,
Misumi:2012eh,Cho:2013yha,Cho:2015ffa,Durr:2017wfi}.
As a different avenue, there is a lattice fermion formulation known as the minimally doubled fermion \cite{Karsten:1981gd,Wilczek:1987kw,Creutz:2007af,Borici:2007kz,Bedaque:2008xs,Bedaque:2008jm,
Capitani:2009yn,Kimura:2009qe,Kimura:2009di,Creutz:2010cz,Capitani:2010nn,Tiburzi:2010bm,Kamata:2011jn,Misumi:2012uu,Misumi:2012ky,Capitani:2013zta,Capitani:2013iha,Misumi:2013maa,Weber:2013tfa,Weber:2017eds,Durr:2020yqa}. With keeping $U(1)$ part of chiral symmetry this setup yields two flavors, which is a minimal number of species allowed by Nielsen-Ninomiya's no-go theorem. The drawback of this setup is the explicit breaking of part of C,P,T and hypercubic symmetry, thus the tuning of parameters is required in the simulation.

Our focus in this work is mainly laid on the four-dimensional Wilson fermion and its novel use.
As well-known, the Wilson term breaks the $U(4)\times U(4)$ flavor-chiral symmetry of naive lattice fermion to $U(1)_{V}$ vector symmetry.
However, the fermion with the parameter condition $m+4 r=0$ (with the mass parameter $m$ and Wilson-fermion parameter $r$) has the enhanced symmetry $U(1)_{V}\times U(1)_{\overline V}$ \cite{Creutz:2011cd, Kimura:2011ik} as the extra $U(1)_{\overline V}$ symmetry is restored on the central one of five branches of the Wilson Dirac spectrum.
This setup is termed as a ``central-branch Wilson fermion'', which is a six-flavor setup and has been investigated in terms of strong-coupling QCD~\cite{Kimura:2011ik}, the Gross-Neveu model~\cite{Creutz:2011cd} and the lattice perturbation~\cite{Misumi:2012eh,Chowdhury:2013ux}.
Its significant property is that the extra $U(1)_{\overline V}$ symmetry prohibits additive mass renormalization. 
Recently, it was shown that this setup is free from the sign problem since the Dirac determinant is positive semi-definite with the central-branch condition $m+4 r=0$ \cite{Misumi:2019jrt}.
Furthermore, the symmetries of the fermion in two dimensions are elucidated, and the $\mathbb{Z}_2$ 't~Hooft anomaly \cite{tHooft:1979rat, Frishman:1980dq,Wen:2013oza, Kapustin:2014lwa, Cho:2014jfa, Wang:2014pma,Witten:2016cio, Tachikawa:2016cha, Gaiotto:2017yup, Tanizaki:2017bam, Shimizu:2017asf, Tanizaki:2017qhf, Sulejmanpasic:2018upi, Tanizaki:2018xto} among $U(1)_{V}\times U(1)_{\overline V}$ symmetry, lattice translation and lattice rotation symmetries is shown to give a restriction on the nonperturbative properties of $U(1)$ gauge theory and Gross-Neveu model with this fermion setup \cite{Misumi:2019jrt}. 
The absence of additive mass renormalization is also understood in terms of the 't Hooft anomaly in two dimensions.

In this paper we investigate properties and varieties of four-dimensional central-branch Wilson fermions.
We first perform comprehensive study on the formulation with introducing several new insights, where we discuss its construction, the prohibition of additive mass renormalization and its interpretation in terms of 't Hooft anomaly, spontaneous symmetry breaking in the strong-coupling QCD, the absence or the solution of the sign problem for quark determinant, and the possibility of its practical use.
The results indicate the usefulness of the six-flavor central-branch fermion for many-flavor QCD simulation.

We then construct several varieties of the central-branch fermions and study their properties with special attention to their symmetries.
Among them, the two-flavor version is constructed by modifying the Wilson hopping term slightly.
The Dirac spectrum of this two-flavor version has seven branches, where two species live at the central branch.
Although the hypercubic symmetry is reduced to its cubic subgroup, the other symmetries including C, P, T, $U(1)_{V}$ and $U(1)_{\overline V}$ are common to those of the original central-branch fermion.
We study the properties of the two-flavor version, including the additive mass renormalization, SSB of parity and $U(1)_{\overline V}$ symmetry in the strong-coupling lattice QCD, the absence of the sign problem and the parameter tuning procedure for Lorentz symmetry restoration.
In lattice perturbation theory, we find that the sum of masses of the two flavors ($m_{u}+m_{d}$) at the central branch is free from renormalization, while their difference ($m_{u}-m_{d}$) suffers from additive renormalization due to the explicit breaking of hypercubic symmetry. We argue that the existence of hypercubic symmetry is essential for the absence of additive mass renormalization for all the six flavors in the original central-branch Wilson fermion.

We also revisit the staggered-Wilson fermion with the two-flavor central branch, which is obtained from the eight-flavor central-branch Wilson fermion via spin diagonalization.
We investigate its extra symmetry and argue that this ``central-branch staggered-Wilson fermion" has a stable central branch without any additive mass renormalization, thus it can be regarded as a minimally doubled fermion with less symmetry breaking.

The structure of this paper is as follows:
In Sec.~\ref{sec:CB} we investigate the original version of the four-dimensional central-branch Wilson fermion.
In Sec.~\ref{sec:2f} we introduce the two-flavor central-branch fermion and discuss its properties.
In Sec.~\ref{sec:Nf} we discuss other types of central-branch fermions.
In Sec.~\ref{sec:SWF} we review the central-branch staggered Wilson fermion and discuss its extra symmetry.
Section \ref{sec:SD} is devoted to a summary and a discussion.


\section{Central-branch Wilson fermion}
\label{sec:CB}

In this section, we revisit and discuss the central-branch Wilson fermion in four dimensions.
We first introduce its lattice action and its flavor-chiral symmetry.
We then move to the properties, including the absence of additive mass renormalization, the symmetry breaking in the strong-coupling QCD and the absence of the sign problem.

\subsection{Wilson fermion and central-branch condition}

\begin{figure}
\centering
\includegraphics[width=8cm]{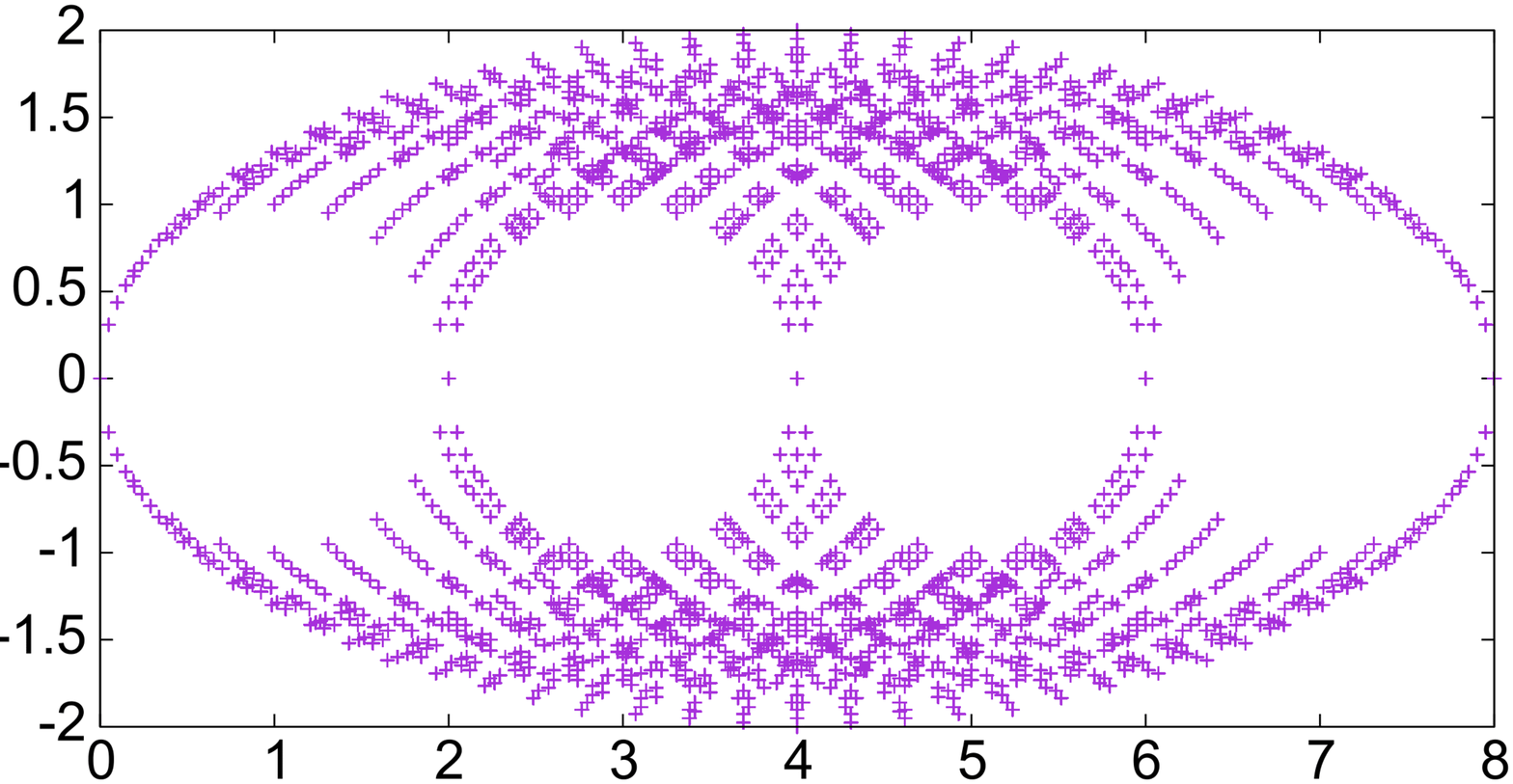} 
\caption{Free Dirac spectrum of Wilson fermion ($r=1$) with $m=0$ on a $20^4$ lattice.
The degenerate spectrum of 16 species for naive fermions are 
split into five branches with 1, 4, 6, 4 and 1 species.}
\label{fig:Wil}
\end{figure}

The lattice Wilson fermion action in four dimensions is 
\begin{equation}
S_{\rm W} =\sum_{n}\sum_{\mu}\overline{\psi}_{n}\gamma_{\mu}D_{\mu}\psi_{n}
\,+\,\sum_{n}m\overline{\psi}_{n}\psi_{n}
+\,r \sum_{n}\sum_{\mu}\overline{\psi}_{n}(1-C_{\mu}){\psi}_{n},
\label{WilS}
\end{equation}
where $D_{\mu}\equiv(T_{+\mu}-T_{-\mu})/2$, 
$C_{\mu}\equiv (T_{+\mu}+T_{-\mu})/2$ with $T_{\pm\mu}\psi_{n}=U_{n,\pm\mu}\psi_{n\pm\mu}$.
In a free theory, we just set $U_{n,\pm\mu}={\bf 1}$.
The sum $\sum_n$ is the summation over lattice sites $n=(n_{1},n_{2},n_{3},n_{4})$. 
A free Wilson fermion has the Dirac spectrum depicted in Fig.~\ref{fig:Wil}.
The degeneracy of 16 species of naive fermion is lifted and they are split into five branches, 
at which 1, 4, 6, 4 and 1 flavors live.

We next study its flavor-chiral symmetry.
We first remind ourselves that the four-dimensional massless naive fermion has $U(4)\times U(4)$ flavor-chiral symmetry \cite{Svetitsky:1980pa,Blairon:1980pk,Kimura:2011ik} in a free theory, which is a subgroup of the whole $U(16)\times U(16)$ symmetry.
These are symmetries under the following transformations,
\bea
\psi_n&\rightarrow& \exp\Big[\im \sum_\A \left(\theta _\A^{(+)}\Gamma^{(+)}_\A+\theta_\A^{(-)}\Gamma^{(-)}_\A\right)\Big]\psi_n\,,\,\,\,\,\,\,\,\, \nonumber\\
\overline{\psi}_n&\rightarrow&\overline{\psi}_n\exp\Big[\im \sum_\A\left(-\theta_\A^{(+)}\Gamma^{(+)}_\A+\theta _\A^{(-)}\Gamma^{(-)}_\A\right)\Big]\,,
\label{sym_naive_all}
\eea
where $\Gamma^{(+)}_\A$ and $\Gamma^{(-)}_\A$ are site-dependent $4\times 4$ matrices,
\begin{eqnarray}
\label{4d:sym_t+}
\Gamma^{(+)}_\A&\in &\left\{\mathbf{1}_4\,,\,\, (-1)^{n_1+\ldots+ n_4}\gamma_5\,,\,\,(-1)^{\check{n}_\mu}\gamma_\mu\,,\,\,(-1)^{n_\mu}i \gamma_\mu\gamma_5
\,,\,\,(-1)^{n_{\mu,\nu}}\frac{i \,[\gamma_\mu\,,\gamma_\nu]}{2}\right\}\,,\\
\label{4d:sym_t-}
\Gamma^{(-)}_\A&\in &\left\{(-1)^{n_1+\ldots+ n_4}\mathbf{1}_4\,,\,\, \gamma_5\,,\,\,(-1)^{n_\mu}\gamma_\mu\,,\,\,(-1)^{\check{n}_\mu}i \gamma_\mu\gamma_5
\,,\,\,(-1)^{\check{n}_{\mu,\nu}}\frac{i \,[\gamma_\mu\,,\gamma_\nu]}{2}\right\}\,,
\end{eqnarray}
with $\check{n}_\mu=\sum _{\rho\neq\mu}n_{\rho}$, $n_{\mu,\nu}=n_\mu+n_\nu$ and $\check{n}_{\mu,\nu}=\sum_{\rho\neq\mu,\nu}n_\rho$.
It is notable that the onsite fermion mass term $\bar{\psi}_{n} \psi_{n}$ breaks this $U(4)\times U(4)$ to the $U(4)$ subgroup $\Gamma^{(+)}_\A$.
In the presence of the Wilson term the $U(4)\times U(4)$ invariance is broken
to the $U(1)$ invariance under $\mathbf{1}_{4}$ in Eq.~(\ref{4d:sym_t+}).

In Refs.~\cite{Creutz:2011cd, Kimura:2011ik}, it was shown that the Wilson fermion with the ``central-branch" condition, 
\be
M_{W}\equiv  m+4r=0,
\label{eq:con}
\ee
has an extra $U(1)$ symmetry, denoted as $U(1)_{\overline{V}}$.
It becomes clear if one is reminded that  
the onsite term ($\sim\bar{\psi}_{n}\psi_{n}$) breaks all the invariance under 
the transformation $\Gamma^{(-)}_{\A}$ in Eq.(\ref{4d:sym_t-}).
Thus, dropping onsite terms can restore some invariance under the group,
and the action comes to have larger symmetry.

\begin{figure}
\begin{center}
\includegraphics[height=6cm]{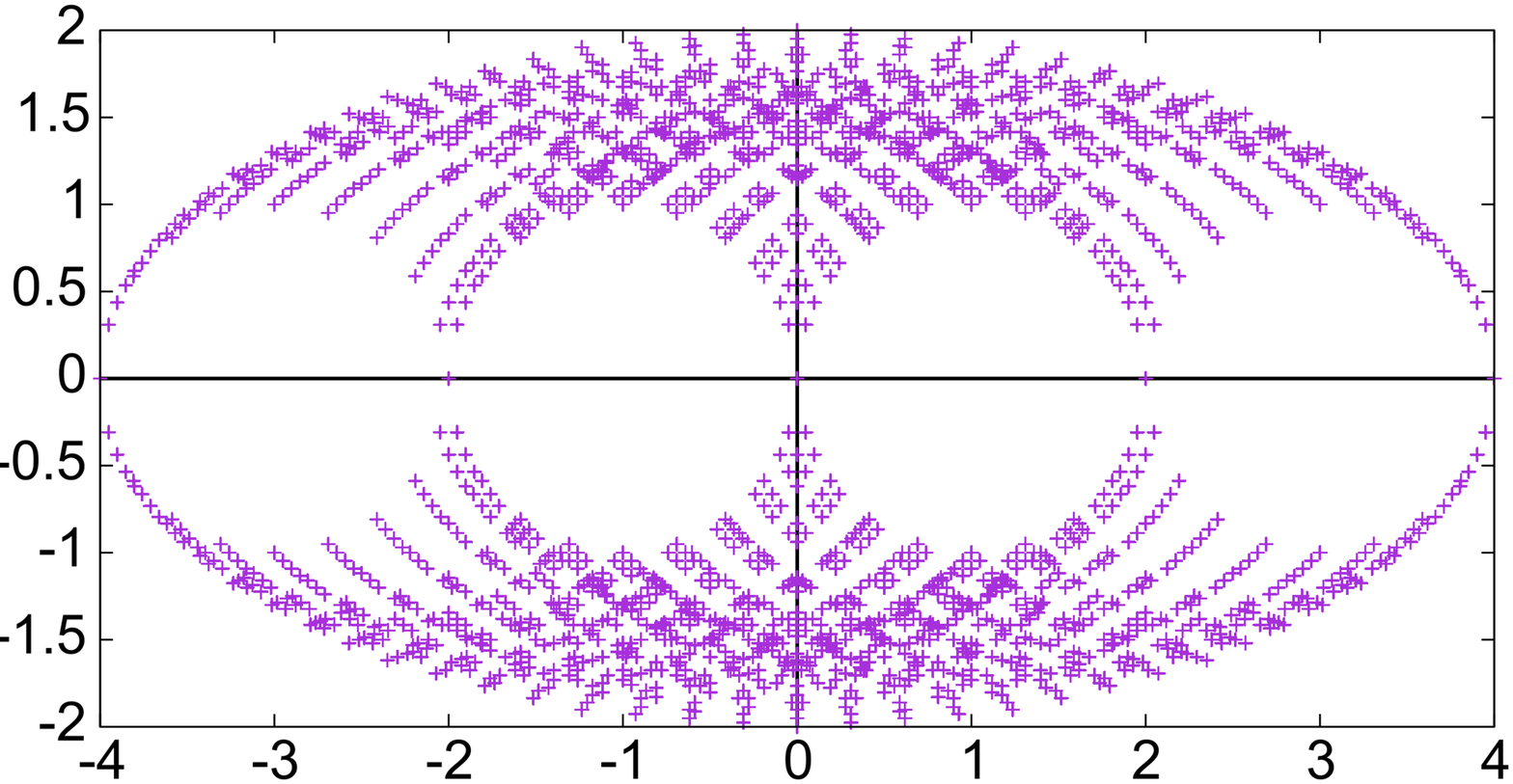} 
\caption{Free Dirac spectrum of Wilson fermion ($r=1)$ with $M_{W}=m+4=0$ on a $20^4$ lattice. 
The central branch with two species crosses the origin. 
}
\label{fig:WilD}
\end{center}
\end{figure}

The free Wilson fermion with this condition (\ref{eq:con}) gives six-flavor massless fermions in the continuum, 
which correspond to the central branch of the Wilson Dirac spectrum as shown in Fig.~\ref{fig:WilD}. 
They are excitations around the Dirac zeros at $p=(\pi,\pi,0,0)$,  $(\pi,0,\pi,0)$,  $(\pi,0,0,\pi)$,  $(0,\pi,\pi,0)$,  $(0,\pi,0,\pi)$ and  $(0,0,\pi,\pi)$ in the momentum space.
This setup is called the ``central-branch Wilson fermion".
Its lattice action is given by
\begin{equation}
S_{\rm CB}=\sum_{n,\mu}\left(\bar{\psi}_{n}\gamma_{\mu}D_{\mu}\psi_{n}-r\bar{\psi}_{n}C_{\mu}\psi_{n}\right),
\label{CB}
\end{equation}
which is invariant under the ordinary $U(1)_V$ transformation generated by $\Gamma^{(+)}=\bm{1}_4$, 
\begin{equation}
U(1)_V: \psi_n \mapsto \rme^{\im \theta}\psi_n,\quad \overline{\psi}_n\mapsto \overline{\psi}_n \rme^{-\im \theta}, 
\label{eq:ordinary_U1}
\end{equation}
and the extra $U(1)$ symmetry generated by $\Gamma^{(-)}=(-1)^{\sum_{\mu}n_{\mu}}\bm{1}_4$, 
\begin{equation}
U(1)_{\overline{V}}: \psi_n\mapsto \rme^{\im (-1)^{\sum_{\mu}n_{\mu}}\theta}\psi_n,\; \overline{\psi}_n\mapsto \overline{\psi}_n \rme^{\im (-1)^{\sum_{\mu}n_{\mu}}\theta}. 
\label{CBsym}
\end{equation}  
It is notable that this extra symmetry prevents the on-site mass term $\bar{\psi}\psi$ from being generated by loop effects, and eventually prohibits additive mass renormalization \cite{Misumi:2012eh} as we will show in the next subsection. 
The other symmetries of this central-branch fermion are the same as those of the usual Wilson fermion, including hypercubic symmetry (lattice rotational symmetry), lattice translation, charge conjugation, parity, $\gamma_{5}$-hermiticity and reflection positivity.

\subsection{Properties of central-branch fermions}
\label{sec:prop}

In this subsection, we study the properties of the central-branch Wilson fermion with introducing several new insights.
We discuss the absence of additive mass renormalization, the condensate and spontaneous symmetry breaking in the strong-coupling limit, the absence or the solution of the sign problem, and possibility of the practical use.


\subsubsection{Absence of additive mass renormalization}

\begin{figure}
\begin{center}
\includegraphics[width=8cm]{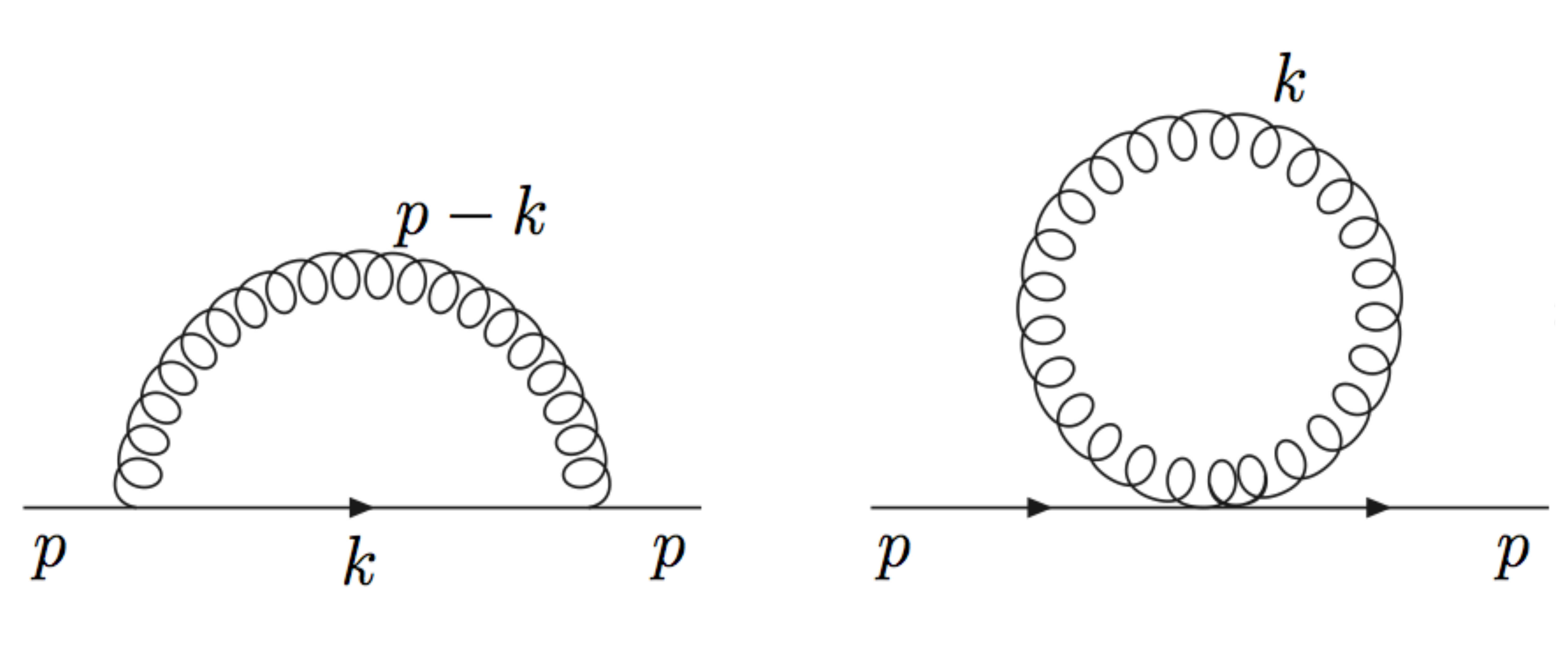} 
\end{center}
\caption{The diagrams contributing to the one-loop quark self-energy. 
The additive mass renormalization from these diagrams are shown to be zero in the original central-branch fermion. }
\label{fig:diag}
\end{figure}

In \cite{Misumi:2012eh,Chowdhury:2013ux} it is shown that the setup is free from the additive quark mass renormalization in the one-loop lattice perturbation theory. 
For simplicity, we take $r=1$ here.
${\mathcal O}(1/a)$ quark self-energy at one-loop level $\Sigma_{0}$ is composed of the sunset $\Sigma_{0}^{(\alpha)}({\rm sun})$ and tadpole $\Sigma_{0}^{\alpha}({\rm tad})$ contributions as shown in Fig.~\ref{fig:diag}.
Here $\alpha=1,2,...,6$ correspond to the six poles of the propagator $\pi_{\mu}^{(1)}=(0,0,\pi,\pi)$, $\pi_{\mu}^{(2)}=(0,\pi,0,\pi),\cdot\cdot\cdot$. 

In lattice QCD with the central-branch Wilson fermion, the gauge propagator with the Feynman gauge $G$, the quark propagator $S$ and the fermion-gauge boson vertex $V$ are given as,
\be
G^{ab}_{\mu\nu}(p) ={4 \delta_{\mu\nu} \delta^{ab}\over{a^{2}}}{1\over{\sum_{\sigma} \sin^{2}{p_{\sigma}\over{2}}}},
\ee
\be
S^{lm}(p) = {a \delta^{lm} \over{ i\sum_{\sigma}\gamma_{\sigma} \sin p_{\sigma}  -r\sum_{\sigma} \cos p_{\sigma}   }},
\ee
\begin{align}
(V^{a})_{\mu}^{mn}(k,p) = -g_{0} (T^{a})^{mn}\left[ i\gamma_{\mu}\cos {k_{\mu}+p_{\mu} \over{2}} 
+r \sin  {k_{\mu}+p_{\mu} \over{2}}  \right],
\end{align}
where $\mu,\nu$ stand for spacetime indices, $a,b$ for Lie group generators and $m,l,n$ for their matrix components. The external momentum is taken to be the values at the Dirac zeros in the following calculations.

The fermion self-energy from the sunset diagram is 
\be
\Sigma^{(\alpha)}({\rm sun}) = \int {d^{4}k \over{(2\pi)^{4}}} \sum_{\mu}G_{\mu\mu}^{ab} (p-k) (V^{b})_{\mu}^{lm}(k,p) S^{mn}(k) (V^{a})_{\mu}^{nl}(p,k).
\ee
Then, $\Sigma_{0}^{(\alpha)}({\rm sun})$ is obtained as
\begin{align}
\Sigma_{0}^{(\alpha)}({\rm sun})&= {g_{0}^{2} C_{F}\over{4a}}\int^{\pi}_{-\pi}{d^{4}k\over{(2\pi)^{4}}}\sum_{\rho} { (\cos^{2}{k_{\rho}\over{2}} - \sin^{2}{k_{\rho}\over{2}})(\sum _{\lambda}\cos k_{\lambda}) 
+\sin^{2} k_{\rho}  
\over{\left(\sum_{\lambda}\sin^{2} {k_{\lambda} + \pi_{\lambda}^{(\alpha)} \over{2}}\right)\left(\sum_{\mu}\sin^{2}k_{\mu}+(\sum _{\mu}\cos k_{\mu})^{2}\right)}}e^{i\pi_{\rho}^{(\alpha)}}\,\,\,\,=\,\,\,0,
\end{align} 
where $e^{i\pi_{\rho}^{(\alpha)}}$ takes $+1$ or $-1$ depending on the direction $\rho$. 
The difference of these signs leads to cancelation between 
dimensions $\rho$ in the integral for any pole $\alpha$.
The cancellation for $r\not=1$ is also verified numerically.

The contribution from the tadpole diagram $\Sigma_{0}^{(\alpha)}({\rm tad})$ is given by 
\begin{equation}
\Sigma_{0}^{(\alpha)}({\rm tad}) =-\int^{\pi}_{-\pi}{d^{4}k\over{(2\pi)^{4}}}
{g_{0}^{2}C_{F}  \over{8a\sum_{\lambda}\sin^{2}{k_{\lambda}\over{2}}}}
\sum_{\rho}\cos \pi^{(\alpha)}_{\rho}\,\,\,=\,\,\,0,
\end{equation}
where cancellation between $\rho$ occurs for any of the six poles. 

We finally obtain
\be
\Sigma_{0}^{(\alpha)}=\Sigma_{0}^{(\alpha)}({\rm sun})+\Sigma_{0}^{(\alpha)}({\rm tad})=0,
\ee
for each of the six poles.
This result indicates the absence of additive renormalization for all the six species at the central branch.
The central branch of Dirac spectrum remains at the origin of the complex plane even if we introduce gauge field.
However, it is somewhat strange that only the restoration of $U(1)_{\overline V}$ symmetry prohibits additive mass renormalization of all the six flavors at the central branch.
We will later find that the existence of hypercubic symmetry as well as $U(1)_{\overline V}$ is essential for this property.

In Ref.~\cite{deForcrand:2012bm}, it is verified in the study of lattice QCD that the additive mass renormalization is absent around the central branch of the staggered-Wilson fermion, which is regarded as the staggered version of the central-branch fermion and will be discussed in detail in Sec.~\ref{sec:SWF}. It implies that one can perform the lattice Monte Carlo simulation without additive mass renormalization by use of the central-branch fermions. 

It is also possible to understand the absence of additive mass renormalization in terms of 't Hooft anomaly matching. First of all, this property means the prohibition of a trivially gapped phase. We consider that it results from a certain mixed 't Hooft anomaly among the symmetries of the systems, including $U(1)_{\overline V}$, hypercubic symmetry, translation invariance, etc.
Indeed, as we will show in the next section, additive mass renormalization is not fully forbidden in the central-branch fermion with explicit breaking of hypercubic symmetry.
We also note that the mixed anomaly among part of hypercubic symmetry, part of translation invariance and $U(1)_{\overline V}$ forbids the trivially gapped phase or the additive mass renormalization in the two-dimensional central-branch Wilson fermion as shown in Ref.~\cite{Misumi:2019jrt}.
More detailed study on the 't Hooft anomaly of the four-dimensional central-branch fermion is left to a future work.


\subsubsection{Symmetry breaking in strong-coupling QCD}

\begin{figure}
\begin{center}
\includegraphics[height=6cm]{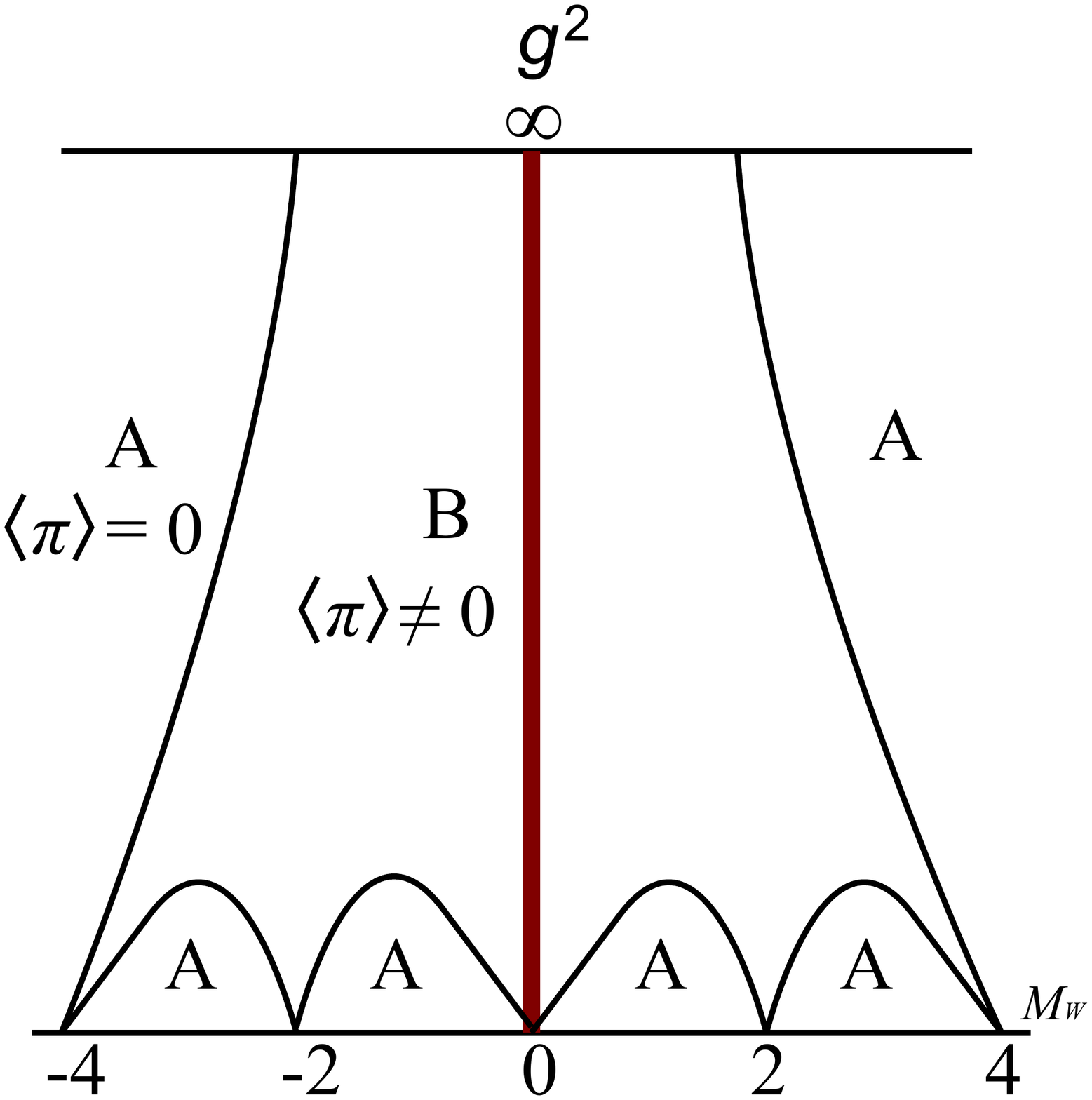} 
\caption{The conjectured Aoki phase diagram for Wilson fermion. The red line corresponds to
the central branch, where the extra symmetry $U(1)_{\overline{V}}$ emerges and is spontaneously broken due to the non-perturbative effect.}
\label{fig:WilA}
\end{center}
\end{figure}

The central branch with the condition $M_{W}=m+4r=0$ in the Wilson fermion corresponds to the central cusp 
in the conjectured parity phase diagram (Aoki phase diagram \cite{Aoki:1983qi,Aoki:1986kt,Aoki:1986xr,Aoki:1987us,Sharpe:1998xm,Creutz:1996bg}) as shown in Fig.~\ref{fig:WilA} and this parameter set is expected to be within the parity broken phase at least in strong and middle gauge coupling regions.
It is unnecessary to take care of the parity breaking when we take the chiral and continuum limits from the parity-symmetric phase.
However, since this setup is free from the sign problem only with the exact central-branch condition,
the question whether we can take the continuum limit from the parity-broken phase is also of importance.

The results on the strong-coupling and large-$N_{c}$ lattice QCD around the central branch \cite{Kimura:2011ik} show that the condensates are given by
\begin{equation}
\sigma = \dfrac{M_W}{4r^2} \,,\,\,\,\,\,\,\,\,\,\,\,\,\,  \pi =\dfrac{1}{16r^4(1+r^2)} (8r^4 - M_W^2(1+r^2)), 
\end{equation}
where $\sigma$ and $\pi$ stand for the chiral and pion condensates
$\langle\bar{\psi}\psi\rangle$, $\langle\bar{\psi}\gamma_{5}\psi\rangle$ respectively.
It is also shown that one of mesonic excitations in the scalar--pseudo-scalar--axial-vector sector has the following mass expression
\begin{eqnarray}
\cosh (m_{SPA}\,) = 1 + \frac{2 M_W^2(16+M_W^2)}{16-15M_W^2}.
\label{dispersion}
\end{eqnarray} 
With the central-branch condition $M_{W}=m+4r=0$, 
we obtain 
\begin{align}
&\sigma=0, \quad\quad \pi=1/2(1+r^2)\,,
\\
&m_{SPA}\,=\,0\,.
\end{align}
This result means that the extra symmetry $U(1)_{\overline V}$ and the parity invariance are spontaneously broken due to the condensate $\langle \bar{\psi}\gamma_{5}\psi\rangle$ instead of $\langle \bar{\psi}\psi\rangle$, and it leads to a massless Nambu-Goldstone boson. 
It is notable that $U(1)_{\overline V}$ emerges only at the central branch $M_{W}=0$ and it is spontaneously broken due to the non-perturbative effect.

The above result of the strong-coupling lattice QCD with the central-branch fermion
($\langle \bar{\psi}\psi\rangle=0,\,\langle \bar{\psi}\gamma_{5}\psi\rangle\not=0$)
indicates that the roles of $\bar{\psi}\psi$ and $\bar{\psi}\gamma_{5}\psi$ are exchanged.
It can be rephrased that the mass basis in this formulation is different from that of the usual lattice QCD.
As discussed in Ref.~\cite{Misumi:2012eh}, we may be able to interpret that the central branch fermion is regarded as an automatic realization of the maximally-twisted-mass Wilson fermion \cite{Sint:2007ug,Shindler:2007vp} since it is regarded as the average of the two edge branches and the central branch is located between them.
Both of the twisted-mass and the six-flavor central-branch fermions are free from ${\mathcal O}(a)$ lattice artifacts since they preserve a subgroup of chiral symmetry. As we will see in the next section, however, the two-flavor central-branch fermion has another source of ${\mathcal O}(a)$ artifacts originating in hypercubic symmetry breaking.

\subsubsection{Sign problem and practical use}

For negative mass $m<0$ in the Wilson and Wilson-like fermions, the semi-positivity of the quark determinant is not guaranteed since there can be an odd number of modes with real-negative eigenvalues.
However, it was proved that the quark determinant of Wilson fermion with the central-branch condition is positive semi-definite on the even-site lattice in Ref.~\cite{Misumi:2019jrt}.

We now extend the proof to four dimensions \cite{Misumi:2019jrt}:
We only consider the case that each number of the lattice sizes $N_{1}, N_{2},N_{3},N_{4}$ in all four dimensions is even-integer. 
We denote the central-branch Dirac operator as 
\be
\mathsf{D}=\sum_{\mu}(\gamma_\mu D_\mu-r C_\mu). 
\ee
Even if we introduce link variables in this operator,
there is $\gamma_5$-hermiticity as 
\be
\gamma_5 \mathsf{D} \gamma_5=\mathsf{D}^\dagger ,
\ee
which leads to the pairing of complex eigenvalues $\lambda, \lambda^*$ in the Dirac spectrum. 
This guarantees that the quark determinant is real or zero for Wilson fermion with any mass parameter.

The central-branch Wilson fermion has further property to restrict the quark determinant.
The $U(1)_{\overline{V}}$ symmetry specific to the central-branch condition can be expressed as
\be
\mathsf{D} (-1)^{\sum_{\mu}n_{\mu}}=- (-1)^{\sum_{\mu}n_{\mu}} \mathsf{D},
\ee
which means the pairing of nonzero eigenvalues $\lambda, -\lambda$ in the Dirac spectrum. 
This property is reflected by the point-symmetric Dirac spectrum of the central branch Wilson fermion.
We now define the hermitian Dirac operator as
\be
H=\gamma_5 \mathsf{D}. 
\ee
The $\gamma_5$-hermiticity of $\mathsf{D}$ guarantees $H^\dagger =H$ and its spectrum should be real. 
The $U(1)_{\overline{V}}$ symmetry is expressed for this operator as 
\be
H(-1)^{\sum_{\mu}n_{\mu}}=-(-1)^{\sum_{\mu}n_{\mu}}H, 
\ee
which leads to the pairing of nonzero eigenvalues $\ve, -\ve$ in the spectrum of $H$.
We here ignore zero eigenvalues for a while and label the spectrum of eigenvalues as 
\be
\{\pm \ve_i\}_{i=1,\ldots,N},
\ee
where $N$ is defined as $N=N_{1}N_{2}N_{3}N_{4}$.
Since $N$ is an even integer, we obtain
\be
\mathrm{det}(\mathsf{D})=\mathrm{det}(H)=\prod_{i=1}^{N}\ve_i (-\ve_i) =(-1)^{N}\prod_{i=1}^{N} \ve_i ^2>0. 
\ee
If the spectrum contains zero eigenvalues, we have $\mathrm{det}(\mathsf{D})=0$. 
Therefore $\mathrm{det}(\mathsf{D})$ is positive semi-definite,
\be
\mathrm{det}(\mathsf{D})\ge 0. 
\ee 
We can rephrase this result in terms of spectrum of $\mathsf{D}$:
When there are real negative eigenvalues,
we simultaneously have genuine zero eigenvalues and the determinant becomes zero.
When there are no real negative eigenvalues,
we have no zero eigenvalues and the determinant becomes nonzero and positive-real.
Thus, the determinant is positive semi-definite for any configuration.

In the Monte Carlo simulation with the central-branch fermion,
there are two possible patterns of its use, both of which have their own advantages and disadvantages:

The first pattern is to generate configurations right on the central branch without mass shift and take a continuum limit.
This method is free from the sign problem of quark determinant,
but the parameter set corresponds to the parity-broken phase.
Although it may correspond to the simulation with the maximally-twisted Wilson fermion as discussed in the previous subsection, we have to introduce the twisted mass to prevent the genuine zero modes from mutilating the simulation \cite{Misumi:2012eh}.
It could be possible, but is not a conventional manner.

The second pattern is to generate configurations with ${\mathcal O}$(1) mass shift from the central branch
and take chiral and continuum limits toward the central branch from the parity-symmetric phase.
This method is free from parity breaking, 
but involves the sign problem of quark determinant.
To investigate this case in detail, let us assume a topological charge of gauge configuration be $Q$.
We now make ${\mathcal O}(1)$ positive mass shift from the central branch.
Then, we have the single flavor with the real-eigenvalue contribution $\sim (-4r/a)^{Q}$ at the left edge branch
and the four flavors with $\sim (-2r/a)^{-4Q}$ at the second branch from the left.
We here use the fact that the chiral charges of modes at left-edge and the second branches are opposite.
Since the other eigenvalues are complex-conjugate-pair or positive-real,
the sign of the determinant is investigated in the following expression
\begin{align}
\mathrm{det}(\mathsf{D})\,\propto\,\left(-{4r\over{a}}\right)^{Q} \cdot \left(-{2r\over{a}}\right)^{-4Q}.
\end{align}
If $Q$ is even-integer (odd-integer), it becomes positive (negative).
Thus, we have the sign problem, where the sign of the quark determinant depends on whether the topological charge of configuration is even or odd.
However, this sign problem is easily bypassed.
We can just quench the sign of the determinant to realize gauge theory with the species at the central branch.
The reason why this simple solution works is as follows:
As an artifact of the present system with mass shift,
the resultant theory becomes gauge theory including the $\theta$ term with $\theta=\pi$.
Quenching the sign of the determinant corresponds to eliminating this $\theta$ term.
Thus, the gauge theory without the $\theta$ term coupled to the central-branch species is realized just by removing the sign of the quark determinant.
This way of bypassing the sign problem is first proposed in the study on the staggered-Wilson fermion in Ref.~\cite{deForcrand:2012bm}, and we just apply it to the central-branch Wilson fermion here.

As we have shown in this section, the six-flavor central-branch Wilson fermion has enough symmetries to prohibit the additive mass renormalization for all the six species and it is free from or is able to bypass the sign problem. Thus, the fermion formulation is a promising formulation for six-flavor or twelve-flavor QCD simulations without parameter-tuning.


\section{Two-flavor central-branch fermion}
\label{sec:2f}

\begin{figure}
\begin{center}
\includegraphics[height=4cm]{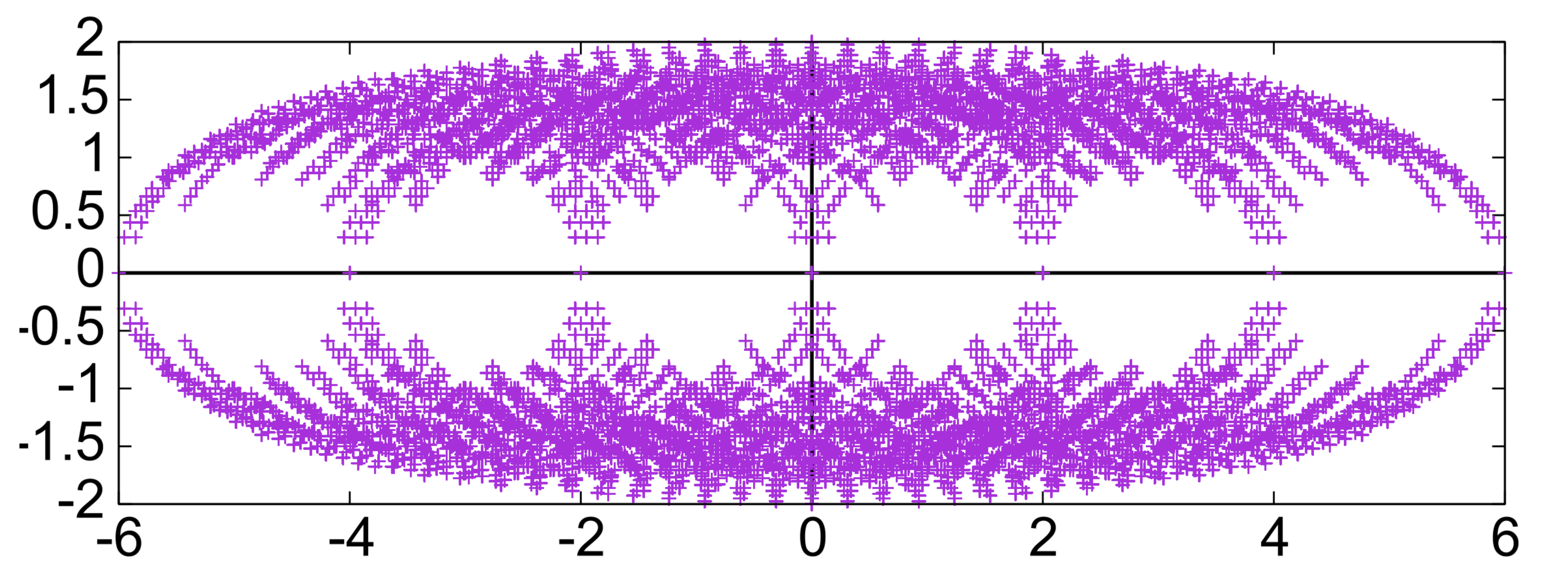} 
\caption{Free Dirac spectrum of the two-flavor central-branch fermion ($r=1$) on a $20^4$ lattice, 
whose hopping term is given by $\sum_{j}C_{j}+3C_{4}$.
1, 3, 3, 2, 3, 3, 1 species live at the seven branches respectively.
}
\label{fig:2c}
\end{center}
\end{figure}

In this section, we construct a two-flavor version of central-branch fermions
and discuss its properties in comparison to the original one.
We first consider a simple modification of hopping terms in the Wilson term as
\begin{equation}
\sum_{\mu=1}^{4} C_{\mu}\,\,\,\to\,\,\, \sum_{j=1}^{3}C_{j}+3C_{4}\,,
\label{mW}
\end{equation}
where we note $C_{\mu}\equiv (T_{+\mu}+T_{-\mu})/2$ with $T_{\pm\mu}\psi_{n}=U_{n,\pm\mu}\psi_{n\pm\mu}$. For a free theory, we set $U_{n,\pm\mu}={\bf 1}$.
Then, the modified Wilson fermion is given as
\be
S=\sum_{n,\mu}\bar{\psi}_{n}\gamma_{\mu}D_{\mu}\psi_{n} +
 \sum_{n} \overline{\psi}_{n}[m+r(6-C_{1}-C_{2}-C_{3}-3C_{4})]\psi_{n}.
\ee
With this central-branch condition $M_{W}=m+6r =0$, 
the action of central-branch Wilson fermion is given by
\be
S_{\rm 2fCB}=\sum_{n,\mu}\bar{\psi}_{n}\gamma_{\mu}D_{\mu}\psi_{n}
-r \sum_{n} \overline{\psi}_{n}(C_{1}+C_{2}+C_{3}+3C_{4})\psi_{n}.
\ee
In a free theory, the Dirac operator in the momentum space is expressed as
\be
\mathsf{D}(p) \,=\, \sum_{\mu=1}^{4} i\gamma_{\mu}\sin p_{\mu}\,-\, r (\sum_{j=1}^{3} \cos p_{j} + 3\cos p_{4})\,.
\ee
The Dirac spectrum for a free theory is depicted in Fig.~\ref{fig:2c}.
The 16-degenerate spectrum is split into seven branches in which 1, 3, 3, 2, 3, 3 and 1 species live.
The two species at the central branch correspond to the two zeros of the Dirac operator
$p=(0,0,0,\pi)$ and $p=(\pi, \pi, \pi, 0)$ in the momentum space.
This fermion action explicitly breaks hypercubic symmetry
into cubic symmetry, while it shares with the original central-branch fermion all the other symmetries and properties, including $U(1)_{V}$, $U(1)_{\overline V}$, C, P, lattice translation, $\gamma_{5}$-hermiticity and reflection positivity.

In the next several subsections, we discuss properties of this two-flavor central-branch fermions in comparison to the original one.
We study the additive mass renormalization in lattice perturbation theory, the parity and $U(1)_{\overline V}$ breaking in the strong-coupling limit, the absence of the sign problem on the central branch, and the necessity of parameter tuning for restoration of Lorentz symmetry.
By comparing the properties of the original and two-flavor central branch fermions, we will find that hypercubic symmetry is essential for the absence of additive mass renormalization of all the species at the central branch.


\subsection{Additive mass renormalization}

As with the original central-branch fermion \cite{Misumi:2012eh,Chowdhury:2013ux},
we can calculate the quark mass renormalization in one-loop lattice perturbation theory.
The quark propagator $S$ and the fermion-gauge boson vertex $V$ are different from those in the original central-branch fermion as
\begin{align}
S^{lm}(p) &= {a \delta^{lm} \over{ i\sum_{\mu}\gamma_{\mu} \sin p_{\mu}  -r\sum_{j=1}^{3} \cos p_{j} -3r \cos p_{4}  }},
\\
(V^{a})_{j}^{mn}(k,p) &= -g_{0} (T^{a})^{mn}\left[ i\gamma_{j}\cos {k_{j}+p_{j} \over{2}} 
+r \sin  {k_{j}+p_{j} \over{2}}  \right],\quad\quad\quad (j=1,2,3)
\\
(V^{a})_{4}^{mn}(k,p) &= -g_{0} (T^{a})^{mn}\left[ i\gamma_{4}\cos {k_{4}+p_{4} \over{2}} 
+3r \sin  {k_{4}+p_{4} \over{2}}  \right],
\end{align}
We now consider the sunset $\Sigma_{0}^{(\alpha)}({\rm sun})$ and tadpole $\Sigma_{0}^{(\alpha)}({\rm tad})$ diagrams. $\alpha=1,2$ correspond to the two zeros denoted as $\pi_{\mu}^{(1)}=(0,0,0,\pi)$ and $\pi_{\mu}^{(2)}=(\pi,\pi,\pi,0)$. 
The external momentum is taken to be the values at the Dirac zeros in the calculation.
$\Sigma_{0}^{(\alpha=1)}({\rm sun})$ is then given by
\begin{align}
\Sigma_{0}^{(\alpha=1)}({\rm sun})= {r g_{0}^{2} C_{F} \over{4a}}\int{d^{4}k\over{(2\pi)^{4}}}
\Big[
&\sum_{j=1}^{3} { ({\mathbf c}_{j}^{2} -r^{2}{\mathbf s}_{j}^{2})(\sum_{i=1}^{3}c_{i} +3c_{4}) 
+s_{j}^{2}  
\over{(\sum_{i=1}^{3}{\mathbf s}_{i}^{2} + {\mathbf c}_{4}^{2})(\sum_{\mu}s_{\mu}^{2} + r^{2}(\sum_{i=1}^{3}c_{i}  +3c_{4})^{2})}}
\nonumber\\
&- { (9r^{2}{\mathbf c}_{4}^{2} -{\mathbf s}_{4}^{2})(\sum_{i=1}^{3}c_{i}  +3c_{4}) 
+3s_{4}^{2}  
\over{(\sum_{i=1}^{3}{\mathbf s}_{i}^{2} + {\mathbf c}_{4}^{2})(\sum_{\mu}s_{\mu}^{2} + r^{2}(\sum_{i=1}^{3}c_{i}  +3c_{4})^{2})}}
\Big],
\end{align} 
where we define $s_{\mu}\equiv \sin k_{\mu}$, $c_{\mu}\equiv \cos k_{\mu}$, ${\mathbf s}_{\mu}\equiv \sin (k_{\mu}/2)$ and ${\mathbf c}_{\mu}\equiv \cos (k_{\mu}/2)$.
There is also the sunset diagram contribution for $\alpha=2$, which is given by
\begin{align}
\Sigma_{0}^{(\alpha=2)}({\rm sun})= {r g_{0}^{2} C_{F} \over{4a}}\int{d^{4}k\over{(2\pi)^{4}}}
\Big[
&\sum_{j=1}^{3} { ({\mathbf s}_{j}^{2} -r^{2}{\mathbf c}_{j}^{2})(\sum_{i=1}^{3}c_{i} +3c_{4}) -s_{j}^{2}  
\over{(\sum_{i=1}^{3}{\mathbf c}_{i}^{2} + {\mathbf s}_{4}^{2} )(\sum_{\mu}s_{\mu}^{2} + r^{2}(\sum_{i=1}^{3}c_{i}+3c_{4})^{2})}}
\nonumber\\
&- { (9r^{2} {\mathbf s}_{4}^{2} -{\mathbf c}_{4}^{2})(\sum_{i=1}^{3}c_{i} +3c_{4}) 
-3s_{4}^{2}  
\over{(\sum_{i=1}^{3}{\mathbf c}_{i}^{2} + {\mathbf s}_{4}^{2} )(\sum_{\mu}s_{\mu}^{2} + r^{2}(\sum_{i=1}^{3}c_{i}+3c_{4})^{2})}}
\Big]\,.
\end{align} 
We can numerically calculate them since they are convergent integrals.
For $r=1$, we obtain
\begin{align}
&\Sigma_{0}^{(\alpha=1)}({\rm sun})\,=\,-0.109985 \, {g_{0}^{2} C_{F} \over{a}}\,,
\\
&\Sigma_{0}^{(\alpha=2)}({\rm sun})\,=\,+0.109985 \, {g_{0}^{2} C_{F} \over{a}}\,.
\end{align}
We also calculate them for other values of $r$. For instance, we obtain $\mp0.0975065$ instead of $\mp0.109985$ in the above equations for $r=0.7$.
We thus conclude $\Sigma_{0}^{(\alpha=1)}({\rm sun}) + \Sigma_{0}^{(\alpha=2)}({\rm sun})=0$.
The contribution from the tadpole diagram to the mass, denoted as $\Sigma^{(\alpha)}_{0}({\rm tad})$, is given by 
\begin{equation}
\Sigma^{(\alpha)}_{0}({\rm tad}) =-{g_{0}^2 C_{F} \over{8a}}\int{d^{4}k\over{(2\pi)^{4}}}
{1  \over{\sum_{\lambda}{\mathbf s}_{\lambda}^{2}}}
\left(\sum_{i}\cos \pi^{(\alpha)}_{i} + 3 \cos \pi^{(\alpha)}_{4} \right)\,\,\,=\,\,\,0.
\end{equation}
It means $\Sigma_{0}^{(\alpha)}({\rm tad})=0$ for either of the two poles. 
Finally, the total contribution to ${\mathcal O}(1/a)$ mass renormalization is given as
\begin{align}
\Sigma_{0}^{(\alpha=1)}\, = -\Sigma_{0}^{(\alpha=2)}\,\not=\,0\,,
\end{align}
This result means that the sum of masses of the two flavors at the central branch is free from additive renormalization, but their difference suffers from it.
It is notable that renormalization of the sum of the masses is prohibited by $U(1)_{\overline V}$, but the breaking of hypercubic symmetry leads to renormalization of the mass difference.

\begin{figure}
\begin{center}
\includegraphics[height=5cm]{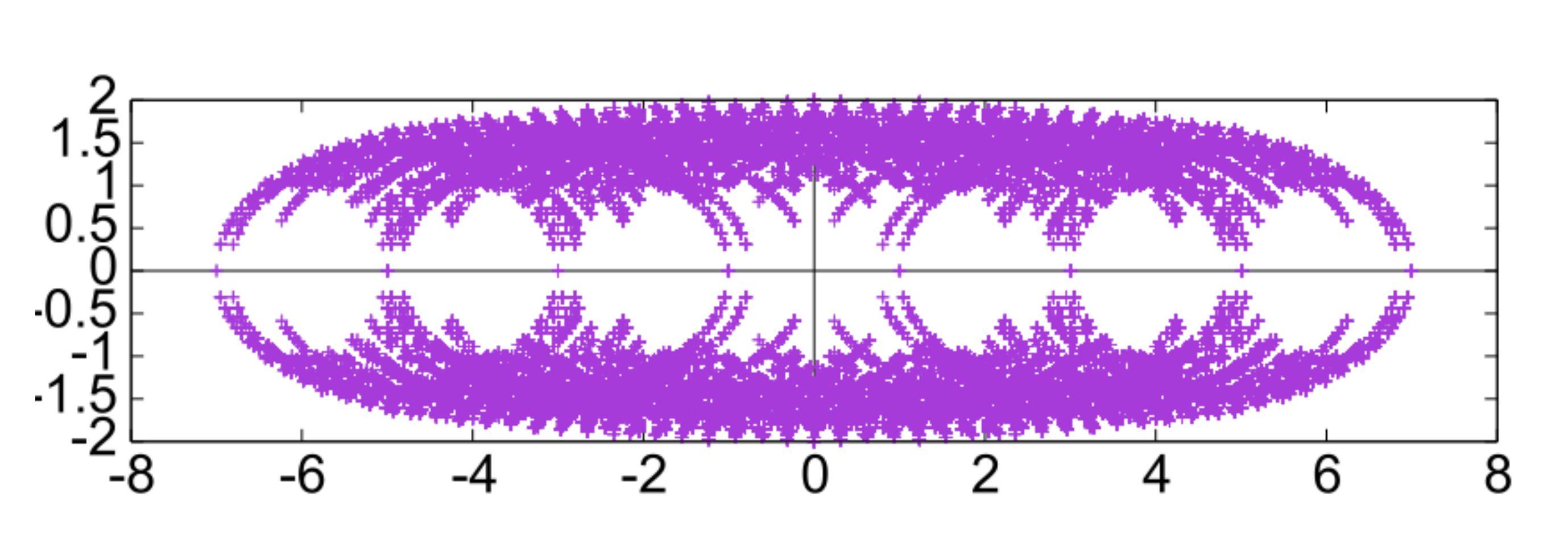} 
\caption{Schematic Dirac spectrum of renormalized two-flavor central-branch fermion, which is $m_{R}=1$ Dirac spectrum in (\ref{eq:renor}).
The central branch is split into two branches.
}
\label{fig:renor}
\end{center}
\end{figure}

The flavor corresponding to the Dirac zero $p=(0,0,0,\pi)$ gets negative mass via loop effects, while the other flavor corresponding to $p=(\pi,\pi,\pi,0)$ gets positive mass.
Roughly speaking, the action of the two-flavor central-branch fermion ($r=1$) is renormalized as
\be
S^{R}=\sum_{n,\mu}\bar{\psi}_{n}\gamma_{\mu}D_{\mu}\psi_{n}
- \sum_{n} \overline{\psi}_{n}(C_{1}+C_{2}+C_{3}+(3+m_{R})C_{4})\psi_{n},
\label{eq:renor}
\ee
where $m_{R}$ stands for a half of the renormalized mass difference.
The Dirac spectrum for the renormalized action is depicted in Fig.~\ref{fig:renor},
where the central branch is split into two branches.
By introducing the counter term $\mu \bar{\psi}_{n} C_{4} \psi_{n}$ with a tuning parameter $\mu$,
we can control this additive renormalization to realize two flavors with arbitrary opposite masses.

This result informs us of important facts on the original central-branch fermion.
The main difference of the two-flavor version from the original one is the breaking of hypercubic symmetry.
It means that the existence of full hypercubic symmetry is essential for the absence of additive mass renormalization of all the six species in the original one.

The additive mass-difference renormalization in the two-flavor version indicates that the symmetries of the system has no mixed 't Hooft anomaly to forbid a trivially gapped phase.
More precisely, gapless fermionic modes in a free theory get massive in an interacting theory and the system is expected to become a trivially gapped phase.
However, since the eight degrees of freedom have negative masses as shown in Fig.~\ref{fig:renor},
the system is speculated in a certain symmetry-protected-topological phase with $U(1)\times U(1)_{\overline V}$.
It could be interesting to study this fermion formulation from this viewpoint.

\subsection{Symmetry breaking in strong-coupling QCD}

In this subsection we take a parallel procedure in the strong-coupling lattice QCD to the original central-branch fermion.
By using hopping operators $P^{\pm}_{\mu}$ and an onsite operator $\hat{M}$, 
a generic lattice fermion action is written as
\begin{equation}
S= \sum_{n,\mu}\bar{\psi}_{n}(P_{\mu}^{+}\psi_{n+\mu}-P_{\mu}^{-}\psi_{n-\mu}) 
+ \sum_{n}\bar{\psi}_{n}\hat{M}\psi_{n}.
\end{equation} 
With these operators an effective action for mesons in the strong-coupling limit is expressed as
\begin{eqnarray}
S_{\rm eff}(\mathcal{M}) &=& N_c \sum_n \left[\sum_\mu {\rm Tr}\, f(\Lambda_{n,\mu}) + {\rm tr} \, \hat{M} \mathcal{M}(n) - {\rm tr} \, \log \mathcal{M}(n) \right] \, ,\\
\Lambda_{n,\mu}&=&\frac{V_{n,\mu} \bar V_{n,\mu}}{N^2_c} , \quad
\mathcal{M}(n)^{\alpha\beta} =  \frac{\sum_a \bar \psi_n^{a,\alpha}\psi_n^{a,\beta}}{N_c}\, ,
\nonumber
\end{eqnarray}
\begin{eqnarray}
V_{n,\mu}^{ab} &=& \bar\psi_n^b P^-_\mu \psi_{n+\hat\mu}^a\, , \quad
\bar V_{n,\mu}^{ab} = -\bar\psi_{n+\hat\mu}^b P^+_\mu \psi_{n}^a\, , \quad \\
{\rm Tr}\, f(\Lambda_{n,\mu})&=& -{\rm tr}\, f\left( - \mathcal{M}(n) \PpT  \mathcal{M}(n+\hat\mu)\PmT
\right)\,, 
\end{eqnarray}
where $N_c$ is the number of colors, ${\rm Tr}$ (${\rm tr}$) is a trace over color (spinor) index, 
and $\mathcal{M}(n)$ is a meson field. $a,b$ are indices for colors, while $\alpha,\beta$ for spinors.
The explicit form of the function $f$ is derived by performing a one-link integral of the gauge field. 
In the large $N_c$ limit, $f(x)$ is
\begin{equation}
f(x) = \sqrt{1+4x}-1-\ln\frac{1+\sqrt{1+4x}}{2} = x + O(x^2)\, .
\label{eq:largeN}
\end{equation}
$f(x)\sim x$ is a good approximation in a large-dimension limit.
For the two-flavor central-branch fermion, we have $\hat M = (m+6r) {\bf
1}_4  = M_{W} \id_{4}$ and
\begin{eqnarray}
P^+_\mu &=& \left\{
\begin{array}{ccc}
 \frac{1}{2} (\gamma_\mu - r) & \mu=1,2,3     \\
 \frac{1}{2}(\gamma_4 - 3r) & \mu=4   \\
\end{array}
\right. ,
 \quad
P^-_\mu =
 \left\{
\begin{array}{ccc}
 \frac{1}{2} (\gamma_\mu + r) & \mu=1,2,3     \\
 \frac{1}{2}(\gamma_4 + 3r) & \mu=4   \\
\end{array}
\right. .
\end{eqnarray}
We here assume a form of meson condensate with
chiral and pion condensates as
\begin{equation}
\mathcal{M}_0 =\sigma{\bf 1}_4 + i \pi \gamma_5.
\end{equation} 
This $\mathcal{M}_0$ is regarded as the vacuum expectation value of $\mathcal{M}(n)$.
Then, the explicit form of the effective action for $\sigma$ and $\pi$ is given by
\begin{align}
S_{\rm eff}&=-4N_{c}{\rm Vol.} \mathcal{V}_{\rm eff}(\sigma,\pi),
\\
\mathcal{V}_{\rm eff}(\sigma,\pi)&=
{1\over{2}}\log (\sigma^{2}+\pi^{2})- M_{W}\sigma
-(1-3r^{2})\sigma^{2}
-(1+3r^{2})\pi^{2}.
\end{align}
We find saddle points of $S_{\rm eff}(\mathcal{M})$ from
\begin{align}
{\delta S_{\rm eff}\over{\delta \sigma}}={\delta S_{\rm eff}\over{\delta \pi}}=0.
\end{align}
Then gap equations are given by
\begin{eqnarray}
(2-6 r^2)\sigma + M_{W} -\frac{\sigma}{\sigma^2+\pi^2} &=& 0 \, ,
\label{g1}
\\
(2+6 r^2)\pi  -\frac{\pi}{\sigma^2+\pi^2} &=& 0 \,.
\label{g2}
\end{eqnarray}
By solving these gap equations we find
\begin{equation}
\sigma = \dfrac{M_W}{12r^2} \,,\,\,\,\,\,\,\,\,\,\,\,\,\,  \pi^2 =\dfrac{1}{144r^4(1+3r^2)} (72r^4 - M_W^2(1+3r^2)),
\label{eq:gapsol} 
\end{equation}
where $\sigma$ and $\pi$ stand for the chiral and pion condensates.
$\pi$ can be nonzero for 
\be
M_{W}^{2}< {72r^4 \over{1+3r^2}}\,,
\ee
which is the Aoki phase region for the present setup in the strong-coupling limit.
With the central-branch condition $M_{W}=0$, we have 
\be
\sigma=0,\quad \pi^2 ={1\over{2(1+3r^2)}}.
\ee
This result indicates that $U(1)_{\overline V}$ is spontaneously broken
due to the parity broken condensate $\langle \bar{\psi}\gamma_{5}\psi\rangle$ instead of $\langle \bar{\psi}\psi\rangle$.

We next look into mass of mesons. 
For this purpose we expand the meson field as
\begin{equation}
\mathcal{M}(n) = \mathcal{M}^\T_0 + \sum_\A \pi^\A(n) \Gamma^\T_\A \,,\quad
\A \in \left\{ S, P, V_\alpha, A_\alpha, T_{\alpha\beta}\right\}\,,
\label{sectors}
\end{equation}
where $S,P, V_{\alpha},A_{\alpha}$ and $T_{\alpha\beta}$ stand for scalar, pseudo-scalar, vector, axial-vector and tensor respectively.
We note
\begin{eqnarray}
\Gamma_S =\frac{{\bf 1}_4}{2}, \ \Gamma_P =\frac{\gamma_5}{2}, \ \Gamma_{V_\alpha}=\frac{\gamma_\alpha}{2},  \ \Gamma_{A_\alpha}=\frac{i \gamma_5\gamma_\alpha}{2}, \
\Gamma_{T_{\alpha\beta}}=\frac{\gamma_\alpha\gamma_\beta}{2i }\ (\alpha < \beta).
\end{eqnarray}
Then the effective action at the second order of $\pi^\A$ is given by
\begin{eqnarray}
S_{\rm eff}^{(2)} &=& N_c \sum_n\biggl[
\frac{1}{2}{\rm tr}\,( \mathcal{M}_0^{-1} \Gamma_\A \mathcal{M}_0^{-1}\Gamma_\B) \,  \pi^\A(n)\pi^\B(n) 
+ \sum_\mu {\rm tr}\, ( \Gamma_\A P^{-}_\mu \Gamma_\B P^{+}_\mu ) \pi^\A(n) \pi^\B(n+\hat\mu)
\biggr] \nonumber \\
&=& N_c\int\frac{d^4 p}{(2\pi)^4} \pi^\A(-p) D_{\A\B}(p) \pi^\B(p) \, ,
\end{eqnarray}
with
\begin{eqnarray}
D_{\A\B} (p) &=& \frac{1}{2} \bigl( \widetilde{D}_{\A\B}(p) + \widetilde{D}_{\B\A} (-p)\bigr),\\[.5ex]
\widetilde{D}_{\A\B}(p) &=& \frac{1}{2}{\rm tr}\,(\mathcal{M}_0^{-1} \Gamma_\A \mathcal{M}_0^{-1}\Gamma_\B)
+ \sum_\mu {\rm tr}\, ( \Gamma_\A P^-_\mu \Gamma_\B P^+_\mu ) e^{i p_\mu} .
\end{eqnarray}
In our case $\mathcal{M}_{0}=\sigma{\bf 1}+i\pi\gamma_{5}$ gives
\begin{equation}
\mathcal{M}_{0}^{-1}={1\over{\sigma^{2}+\pi^{2}}}(\sigma{\bf 1}-i\pi\gamma_{5}).
\end{equation} 
For simplicity, we take $r^{2}=1$.
We now write the inverse meson propagator matrix 
in the $S$-$P$-$A$ sector as
\begin{align}
D_{SPA}=
\left[
\begin{array}{cccccc}
D_{S} & -C & &  &  &   \\
-C & D_{P} & -3s_{4}/2 & -s_{3}/2 & -s_{2}/2  & -s_{1}/2  \\
& 3s_{4}/2 & D_{A_{4}} & & &  \\
& s_{3}/2 & & D_{A_{3}} & & \\
& s_{2}/2 & & & D_{A_{2}} &  \\
& s_{1}/2 & & &  & D_{A_{1}}  \\
\end{array}
\right] \, ,
\end{align}
where components are given by
\begin{align}
D_{S}&= {\sigma^{2}-\pi^{2}\over{2(\sigma^{2}+\pi^{2})^{2}}} -2c_{4}, 
\label{Ds}
\\
D_{P}&= {\sigma^{2}-\pi^{2}\over{2(\sigma^{2}+\pi^{2})^{2}}}
- {1\over{2}}[ c_{1}+c_{2}+c_{3} + 5c_{4}], \\
C&= {i\sigma\pi \over{(\sigma^{2}+\pi^{2})^{2}}},
\\
D_{A_{4}}&= {1\over{2(\sigma^{2}+\pi^{2})}}
- {5c_{4}\over{2}}, \\
D_{A_{3}}&= {1\over{2(\sigma^{2}+\pi^{2})}}
- {c_{3}\over{2}} - 2c_{4}, \\
D_{A_{2}}&={1\over{2(\sigma^{2}+\pi^{2})}}
- {c_{2}\over{2}} -2c_{4}, \\
D_{A_{1}}&={1\over{2(\sigma^{2}+\pi^{2})}}
- {c_{1}\over{2}} -2c_{4},
\label{Da4}
\end{align}
with $s_k =\sin p_k$ and $c_k=\cos p_k$.
\begin{figure}[t]
\begin{center}
\includegraphics[height=9cm]{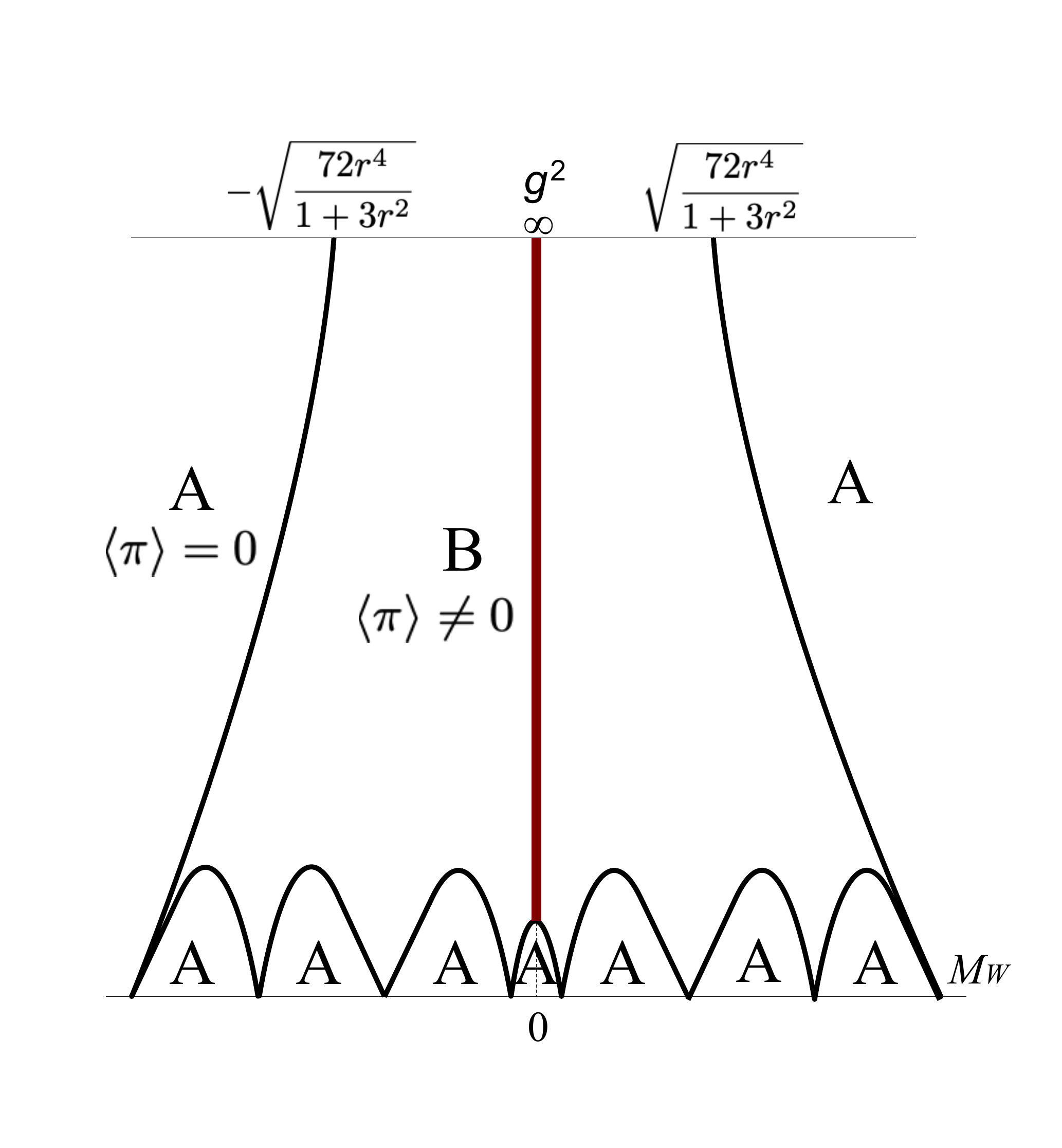} 
\caption{The conjectured Aoki phase diagram for two-flavor central branch fermion. The red line corresponds to the central branch, where the extra symmetry $U(1)_{\overline{V}}$ emerges and is spontaneously broken.
The central cusp is split into two cusps due to the additive renormalization of the mass difference of the two species.}
\label{fig:WilA-2f}
\end{center}
\end{figure}
By diagonalizing this matrix, we can derive 
an explicit form of physical meson propagators in the sector.
We now check that one of meson masses becomes zero at the central branch.
For this purpose we substitute $p=(0,0,0, i m_{\rm m})+(\pi,\pi,\pi,\pi)$ into the propagator.
Then, the $S$-$P$-$A_4$ sector still has off-diagonal components and
what we have to consider is an equation ${\rm det} (D_{SPA_{4}})=0$.
It is expressed as
\begin{align}
D_{S}D_{P}D_{A_4} + {9\over{4}}D_{S}s_{4}^{2} -C^2 D_{A_4}=0.
\end{align}
By substituting the solutions Eq.~(\ref{eq:gapsol}) into the gap equations,
we solve this equation around the central branch $M_{W}\sim0$ and find that one of solutions is given by
\be
\cosh m_{SPA}\,=\, 1+0.737589 M_{W}^2 +{\mathcal O}(M_{W}^4).
\ee
This result means that we have a massless meson for $M_{W}=0$,
\be
\cosh m_{SPA} \,=\, 1 \quad\quad\to\quad\quad m_{SPA}=0.
\ee
This massless meson corresponds to a massless Nambu-Goldstone boson associated with SSB of 
$U(1)_{\overline V}$ symmetry at the central branch.
It indicates that, in the strong-coupling limit, the $U(1)_{\overline V}$ symmetry and the parity invariance are spontaneously broken due to the special condensate.

Since the mass difference between the two flavors on the central branch is additively renormalized as shown in the previous subsection, the above analysis, which is valid just in the strong-coupling region, cannot inform us of the detailed phase structure in the weak-coupling region.
In Fig.~\ref{fig:WilA-2f}, we have shown a conjecture of the parity phase diagram of the two-flavor central-branch fermion.
In the conjecture, the central cusp is split into two cusps due to the additive renormalization of the mass difference.
By tuning the parameter $\mu$ for the counter term $\mu \bar{\psi}C_{4}\psi_{n}$, we control this splitting of the central cusp and realize the system with the two flavors with opposite masses.
It is a nontrivial question whether or not the $U(1)_{\overline V}$ symmetry is spontaneously broken even in the weak-coupling region.

\subsection{Sign problem and practical use}

All the procedure in the proof of the absence of the sign problem is the same as that for the original central-branch fermion.
The $U(1)_{\overline{V}}$ symmetry means 
$H(-)^{\sum_{\mu}n_{\mu}}=-(-)^{\sum_{\mu}n_{\mu}}H$
with the hermitian Dirac operator $H=\gamma_5 \mathsf{D}$. 
It results in the pairing of nonzero eigenvalues $\ve, -\ve$ in the spectrum of $H$ 
as $\{\pm \ve_i\}_{i=1,\ldots,N}$ with $N=N_{1}N_{2}N_{3}N_{4}$.
By taking $N$ as an even integer, we obtain
\be
\mathrm{det}(\mathsf{D})=\mathrm{det}(H)\ge 0. 
\ee

In the Monte Carlo simulation with the two-flavor central-branch fermion,
we have to take account of the fact that the mass difference of the central-branch two flavors suffers additive renormalization.
From the conjectured parity phase diagram in Fig.~\ref{fig:WilA-2f}, the central-branch condition $M_{W}=m+6r=0$ corresponds to the parity-symmetric phase in the weak-coupling region.
As long as we keep this condition, the formulation is free from the sign problem
since the mass-difference renormalization eliminates genuine zero modes for this case.

Let us look into this fact in detail:
We consider the typical case, where the renormalized mass difference is given by $2m_{R}=2$ as with Fig.~\ref{fig:renor}.
We also assume that a topological charge of gauge configuration is $Q$.
We then have the single flavor with the real-eigenvalue contribution $\sim (-7r/a)^{Q}$ at the left edge branch, 
the three flavors with $\sim (-5r/a)^{-3Q}$ at the second branch from the left, the other three flavors with $\sim (-3r/a)^{3Q}$ at the third branch from the left and the other single flavor with $\sim (-r/a)^{-Q}$ at one of the split central branches.
We note that there is no genuine zero mode in this case unless we fine-tune a parameter to cancel out the renormalized mass difference.
We then find that the Dirac determinant
\be
\mathrm{det}(\mathsf{D}) \,\propto\,  \left(-{7r\over{a}}\right)^{Q} \cdot \left(-{5r\over{a}}\right)^{-3Q} \cdot  \left(-{3r\over{a}}\right)^{3Q} \cdot  \left(-{r\over{a}}\right)^{-Q} > 0\,,
\ee
is positive-definite. 
We thus conclude that the formulation is free from the sign problem.


\subsection{Parameter-tuning procedure}

As we have discussed, we need to tune one parameter to control the mass difference of the two flavors in the two-flavor central-branch fermion formulation. 
For this purpose, we tune a parameter for the dimension-3 operator
\be
\bar{\psi}_{n} C_{4} \psi_{n}\,.
\ee
By tuning this parameter, we can realize the two-flavor system with arbitrary opposite masses.

To discuss tuning procedure required for Euclidean Lorentz symmetry restoration in the continuum limit,
we first consider possible operators generated by loop effects.
The discrete symmetries of the system prohibit further emergence of the dimension-$3$ relevant operators
such as $\bar{\psi} \gamma_{4}\psi$.
The dimension-$4$ marginal operators can emerge through loop effects due to the breaking of hypercubic symmetry, but most of them are prohibited by the other discrete symmetries.
The only dimension-$4$ operators we have to care are
\begin{align}
&\bar{\psi}\gamma_{4}\partial_{4}\psi\,,
\\
&\sum_{j=1}^{3} F_{j4}^2 \,,
\end{align}
with $j=1,2,3$.
A coefficient of the former operator is renormalized differently from that of the other dimension-$4$ operators $\bar{\psi}\gamma_{j}\partial_{j}\psi$, while a coefficient of the latter operator is renormalized differently
from that of $F_{ij}^{2}$ with $i,j=1,2,3$.
In other words, the speed of light is renormalized in a unphysical manner in this system
both for quark and gauge fields.
Thus, we have to tune the two marginal parameters to restore the Euclidean Lorentz symmetry.
However, it is worth noting that the tuning procedure for these two parameters is well investigated in the QCD simulation on anisotropic lattices \cite{Morrin:2006tf,Lin:2008pr} and it may be applied to the present case.

As a summary of this section, we make several comments. 
The two-flavor central-branch fermion requires three-parameter tuning for the practical use in lattice QCD. 
Its advantages such as $U(1)_{\overline V}$ symmetry, minimal-doubling and ultra-locality seems to be completely beaten by the drawback.  
However, this disadvantage rather sets off the original central-branch Wilson fermion
since it has no necessity of parameter-tuning in six-flavor lattice QCD.
As we have discussed, this difference originates in the existence of full hypercubic symmetry.
The study of the two-flavor central-branch fermion gives a good lesson that we have to take care of not only lattice flavor-chiral symmetries but also hypercubic symmetry in the central-branch fermions.


\section{Other central-branch Wilson fermions}
\label{sec:Nf}

\begin{figure}
\begin{center}
\includegraphics[height=6cm]{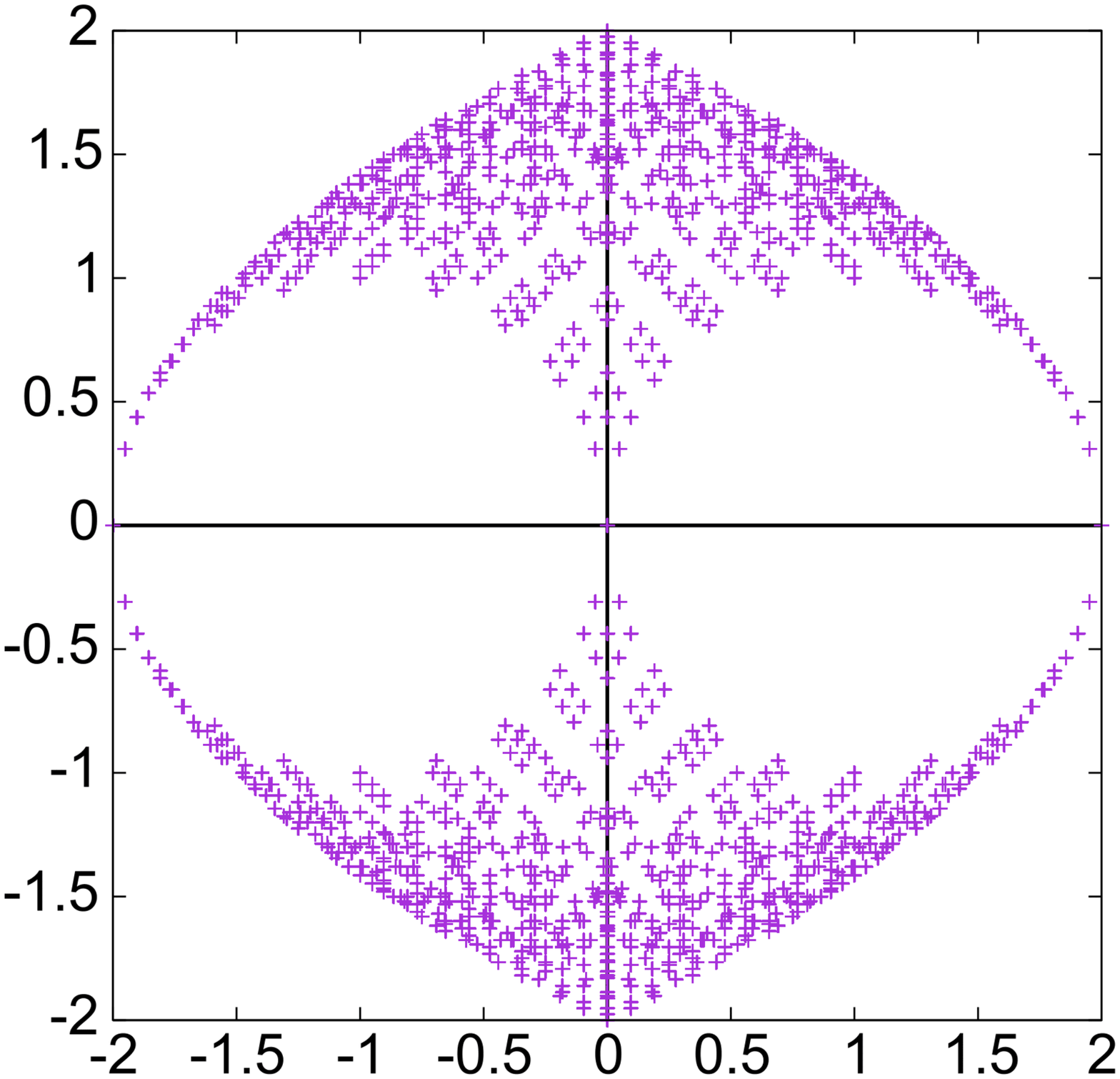} 
\caption{Free Dirac spectrum of the four-dimensional eight-flavor central branch fermion ($r=1$) on a $20^4$ lattice, whose Wilson hopping term is $C_{1}C_{2}+C_{3}C_{4}$.
4, 8 and 4 species live at the three branches respectively.
}
\label{fig:8f}
\end{center}
\end{figure}

In this section, we consider other varieties of central-branch fermions.
For instance, we obtain an eight-flavor central-branch fermion by modification of hopping terms in the Wilson term as
\begin{equation}
\sum_{\mu=1}^{4} C_{\mu}\,\,\,\to\,\,\, C_{12}+C_{34},
\end{equation}
with 
\be
C_{\mu\nu}\equiv \frac{C_{\mu}C_{\nu}+C_{\nu}C_{\mu}}{2}\,.
\ee
With this modification the action of central-branch fermion is given by
\be
S_{\rm 8fCB}=\sum_{n,\mu}\bar{\psi}_{n}\gamma_{\mu}D_{\mu}\psi_{n}
-r \sum_{n} \overline{\psi}_{n}(C_{12}+C_{34})\psi_{n}.
\label{eq:8f}
\ee
This setup corresponds to the central branch of one of the flavored-mass fermions, called the tensor-type fermion \cite{Creutz:2010bm}.
In a free theory, the Dirac operator in the momentum space is expressed as
\be
\mathsf{D}(p) \,=\, \sum_{\mu=1}^{4} i\gamma_{\mu}\sin p_{\mu}\,-\, r ( \cos p_{1} \cos p_{2} + \cos p_{3} \cos p_{4})\,.
\ee
The Dirac spectrum for a free theory with $r=1$ is depicted in Fig.~\ref{fig:8f}.
The 16 species are split into three branches in which 4, 8 and 4 species live.
The eight species at the central branch correspond to the eight zeros of the Dirac operator
$p=(0,0,0,\pi)$, $(0,0,\pi,0)$, $(0,\pi,0,0)$, $(\pi,0,0,0)$, $(\pi, \pi, \pi, 0)$, $(\pi,\pi,0,\pi)$, $(\pi,0,\pi,\pi)$, 
$(0,\pi,\pi,\pi)$ in the momentum space.

Among the flavor-chiral symmetries of the naive fermion,
this setup keeps a relatively large subgroup as
\begin{eqnarray}
\Gamma^{(+)}_\A&\in &\left\{\mathbf{1}_4\,,\,\, (-1)^{n_1+\ldots+ n_4}\gamma_5\,,
\,\,(-1)^{n_{1,2}}\frac{i \,[\gamma_1\,,\gamma_2]}{2}\,,
\,\,(-1)^{n_{3,4}}\frac{i \,[\gamma_3\,,\gamma_4]}{2}\right\}\,,\\
\Gamma^{(-)}_\A&\in &\left\{
(-1)^{\check{n}_{1,3}}\frac{i \,[\gamma_1\,,\gamma_3]}{2}
\,,\,\,(-1)^{\check{n}_{2,4}}\frac{i \,[\gamma_2\,,\gamma_4]}{2}\,,
\,\,
(-1)^{\check{n}_{1,4}}\frac{i \,[\gamma_1\,,\gamma_4]}{2}
\,,\,\,(-1)^{\check{n}_{2,3}}\frac{i \,[\gamma_2\,,\gamma_3]}{2}
\right\}\,.
\end{eqnarray}
It also shares the symmetries and properties including
lattice translation, $\gamma_{5}$-hermiticity, C, P and reflection positivity 
with the original central-branch fermion.
The breaking of hypercubic symmetry is much less severe than that of the two-flavor central-branch fermion.
Regarding restoration of Euclidean Lorentz symmetry in the continuum,
we need parameter-tuning in the gauge-boson part, where the coefficient of $F_{12}^2 + F_{34}^{2}$
is renormalized differently from that of $F_{13}^{2} + F_{23}^{2} + F_{14}^{2} + F_{24}^{2}$.
We also note that the sign problem on the central branch is absent in this case too.

Since the onsite mass term is not invariant under the above flavor-chiral transformations,
the renormalization of the onsite mass term is prohibited.
Furthermore, the absence of additive mass renormalization for each of the eight species is expected since all the possible mass terms for the species seem to be prohibited by the residual hypercubic symmetry and the flavor-chiral symmetry.
It should be verified in future study.

We can also construct another eight-flavor version of central-branch fermions by the modification of the Wilson term $\sum_{\mu=1}^{4} C_{\mu} \to C_{123} + C_{4}$ with
$C_{\mu\nu\rho}\equiv {1\over 6}{ \sum_{\rm perm.}}C_{\mu}C_{\nu}C_{\rho}$.
Although the 16 species are again split into three branches with 4, 8 and 4 species,
the breaking of hypercubic symmetry is severer than the previous version,
thus we speculate that the tuning procedure is required for more parameters.

We consider that there are lots of varieties of central-branch fermions 
and future works will be devoted to their full classification. 
The two-flavor central-branch fermion in five dimensions is briefly addressed in Appendix.~\ref{sec:5d}.


\section{Central branch of Staggered-Wilson fermions}
\label{sec:SWF}

In this section we focus on the staggered fermion \cite{Kogut:1974ag, Susskind:1976jm} and its flavored-mass terms. The argument in this section is in part presented in the proceedings of the lattice conference \cite{Misumi:2012eh} by one of the present author.

Let me start with the action of the staggered fermion,
\begin{align}  
S\,=\,\sum_{xy}\bar{\chi}_{x}[\eta_{\mu}D_{\mu} + m]_{xy}\chi_{y}\,,
\end{align}
where $\chi_{x}$ is an one-component fermion field, and we define $(\eta_{\mu})_{xy}\equiv(-1)^{x_{1}+...+x_{\mu-1}}\delta_{x,y}$ and $D_{\mu}\equiv {1\over{2}}(T_{\mu}-T_{-\mu})$ with $(T_{\pm \mu})_{xy}=U_{x,\pm \mu}\delta_{x\pm \mu,y}$.
$m=m\delta_{x,y}$ is a mass parameter.
This action is obtained from the naive fermion action via the procedure called ``spin diagonalization"
and contains four species called ``tastes".
For simplicity we denote four-dimensional lattice sites as $x$ or $y$ for staggered fermions.
The relevant symmetry of staggered fermion \cite{Golterman:1984cy,Golterman:1985dz,Kilcup:1986dg} is 
\begin{equation}
\{ C_{0}, \,\Xi_{\mu},\, I_{s},\, R_{\mu\nu} \}\,\,\times\,\,\{U^{\epsilon}(1)\}_{m=0},
\label{sSym}
\end{equation}
where $C_{0}$ is staggered charge conjugation, 
$\Xi_{\mu}$ is shift transformation, 
$I_{s}$ is spatial inversion, 
$R_{\mu\nu}$ is hypercubic rotation, 
and $U^{\epsilon}(1)$ is the residual chiral symmetry $\chi_{x}\,\to\, e^{i \theta\epsilon_{x}}\chi_{x}$
with $\epsilon_{x}=(-1)^{\sum_{\mu}x_{\mu}}$, which is expressed as $\gamma_{5}\otimes\xi_{5}$
in the spin-taste representation ($\xi_{5}$ stands for $\gamma_{5}$ in the taste space).
The combinations of these symmetries give physical symmetries, including charge conjugation, parity and spacetime hypercubic symmetry.
The details of symmetries are summarized in App.~\ref{sec:sym}.

\subsection{Staggered-Wilson fermion}

The species-splitting mass term, namely the flavored-mass term, is also introduced into staggered fermions  \cite{Golterman:1984cy, Adams:2009eb,Adams:2010gx,Hoelbling:2010jw}.
They split four degenerate tastes into multiple branches with satisfying other basic properties including $\gamma_{5}$ hermiticity (precisely speaking, $\epsilon_{x}\sim \gamma_{5}\otimes \xi_{5}$ hermiticity). 
In spin-taste representation there are only two types of flavored-mass terms satisfying the $\gamma_{5}$ hermiticity,
corresponding to $\id\otimes\xi_{5}$ and $\id\otimes \sigma_{\mu\nu}$.
These terms are realized as four- and two-hopping terms in
the one-component staggered action up to $\mathcal{O}(a)$ errors.

The four-hopping flavored-mass term \cite{Golterman:1984cy, Adams:2009eb} is given by
\begin{equation}  
M_{\rm A}= \epsilon\sum_{sym} \eta_{1}\eta_{2}\eta_{3}\eta_{4}
C_{1}C_{2}C_{3}C_{4}
= (\id \otimes \xi_{5}) + \mathcal{O}(a)\,,
\label{AdamsM}
\end{equation}
with $(\epsilon)_{xy}=(-1)^{x_{1}+...+x_{4}}\delta_{x,y}$ and
$C_{\mu}=(T_{\mu}+T_{-\mu}^{\dag})/2$.
Here we hide the factor $1/24$ in the symmetric sum $\sum_{sym.}$. 
With this flavored-mass term, the four tastes (species) fall into 
the $\xi_{5}=+1$ two-taste subspace and the $\xi_{5}=-1$ two-taste subspace. 
As a consequence, the corresponding Dirac spectrum has two branches \cite{Adams:2010gx, deForcrand:2011ak}.
By introducing a mass parameter $m=m\delta_{x,y}$ and a Wilson parameter $r=r\delta_{x,y}$ as with the Wilson fermion, the four-hopping staggered-Wilson fermion is expressed as
\begin{equation}  
S_{\rm A}\,=\, \sum_{xy}\bar{\chi}_{x}(D_{\rm A})_{xy}\chi_{y}\,=\,\sum_{xy}\bar{\chi}_{x}[\eta_{\mu}D_{\mu}
+r(1+M_{\rm A})+m]_{xy}\chi_{y}\,.
\label{AdamsS1}
\end{equation}
We note that (\ref{AdamsM}) is derived from the four-hopping flavored-mass term for naive fermions which split sixteen species into two eight-species branches \cite{Creutz:2011cd,M1,Misumi:2012sp,Misumi:2012eh}.
It is schematically expressed as
\begin{equation}
\bar{\psi}_{x}[C_{1}C_{2}C_{3}C_{4}]_{xy}\psi_{y}\,\,\,\,\,\,\,\,\,\,\to\,\,\,\,\,\,\,\,\,\,
\pm\bar{\chi}_{x}[\epsilon\eta_{1}\eta_{2}\eta_{3}\eta_{4}C_{1}C_{2}C_{3}C_{4}]_{xy}\chi_{y}\,.
\label{SD1}
\end{equation}

The two-hopping flavored-mass term \cite{Hoelbling:2010jw} is given by
\begin{equation}  
M_{\rm H}=i(\eta_{12}C_{12}+\eta_{34}C_{34})
= [\id \otimes (\sigma_{12}+\sigma_{34}) ] + \mathcal{O}(a)\, ,
\label{HoelM}
\end{equation}
with
$(\eta_{\mu\nu})_{xy}=\epsilon_{\mu\nu}\eta_{\mu}\eta_{\nu}\delta_{x,y}$,
$(\epsilon_{\mu\nu})_{xy}=(-1)^{x_{\mu}+x_{\nu}}\delta_{x,y}$,
$C_{\mu\nu}=(C_{\mu}C_{\nu}+C_{\nu}C_{\mu})/2$.
This flavored mass splits four tastes into three branches, including one-flavor, two-flavor and the other one-flavor branches. 
By introducing a mass parameter and a Wilson parameter, 
the two-hopping staggered-Wilson fermion is
\begin{align}  
S_{\rm H}\,=\, \sum_{xy}\bar{\chi}_{x}(D_{\rm H})_{xy}\chi_{y}\,=\,
\sum_{xy}\bar{\chi}_{x}[\eta_{\mu}D_{\mu}
+r(2+M_{\rm H})+m]_{xy}\chi_{y}\,.
\label{HoelS}
\end{align}
Eq.~(\ref{HoelM}) is derived from the two-hopping flavored-mass term in Eq.~(\ref{eq:8f}) for naive fermions which split sixteen species into three branches, including four-species, eight-species and the other four-species branches \cite{Creutz:2011cd,M1,Misumi:2012sp,Misumi:2012eh}.
It is expressed as
\begin{equation}
\bar{\psi}_{x}[C_{12}+C_{34}]_{xy}\psi_{y}\,\,\,\,\,\,\,\,\,\,\to\,\,\,\,\,\,\,\,\,\,
\pm\bar{\chi}_{x}[i(\eta_{12}C_{12}+\eta_{34}C_{34})]_{xy}\chi_{y}.
\label{SD2}
\end{equation}

The properties of these staggered-Wilson fermions have been studied in terms of index theorem \cite{Adams:2009eb}, overlap kernel \cite{Adams:2010gx,deForcrand:2011ak}, symmetries \cite{Steve,Misumi:2012sp,Misumi:2012eh}, numerical costs \cite{deForcrand:2011ak,deForcrand:2012bm}, parity phase structure \cite{Creutz:2011cd,Misumi:2011su,Misumi:2012sp}, 
taste-breaking and hadron spectrum \cite{Steve,Misumi:2012sp,Misumi:2012eh}.
We here concentrate on their symmetries in order to study the central-branch staggered-Wilson fermions.
The four-hopping flavored-mass in Eq.~(\ref{AdamsM}) breaks the staggered symmetry in Eq.~(\ref{sSym}) to
\begin{equation}
\{ C_{0}, \Xi_{\mu}', R_{\mu\nu} \}\,,
\label{A1Sym}
\end{equation}
where we define $\Xi_{\mu}'\,\equiv\, \Xi_{\mu}I_{\mu}$. 
Since the action is invariant under the transformation 
$\Xi_{4}I_{s}\sim(\gamma_{4}\otimes 1)$, the physical parity invariance P remains.
Furthermore, $C_{0}$ is also unbroken in this case,
therefore the physical charge conjugation C at the two-flavor branch can be formed in a similar way
to the staggered fermions.
Regarding Euclidean Lorentz symmetry, a combination of the staggered rotation $R_{\mu\nu}$ 
and the shifted-axis reversal $\Xi_{\mu}'$ forms the hypercubic group as with the staggered fermion. 
These facts indicate that the four-hopping staggered-Wilson action (\ref{AdamsS1}) possesses enough discrete symmetries for a correct continuum limit.

On the other hand, the symmetry of the two-hopping staggered fermion in (\ref{HoelS}) is smaller than that of the four-hopping one, which is given by
\be
\{ C_{T}, \Xi_{\mu}', R_{12}, R_{34}, R_{24}R_{31} \}\,.
\ee
Although $C_{0}$ is broken in this action, it is invariant under another special charge conjugation $C_{T}\equiv R_{21}R_{13}^{2}C_{0}$ \cite{Misumi:2012sp,Misumi:2012eh}.
Due to $C_{T}$ and $\Xi_{\mu}'$, the invariances under physical parity and physical charge conjugation are guaranteed at each of the three branches.
However, the breaking of the staggered rotation symmetry leads to the necessity of one-parameter tuning 
to restore Lorentz symmetry, where the coefficient of $F_{12}^2 + F_{34}^{2}$
is renormalized differently from that of $F_{13}^{2} + F_{23}^{2} + F_{14}^{2} + F_{24}^{2}$ \cite{Steve}.
It is a consequence of the fact that the two-hopping staggered-Wilson fermion is derived from the flavored-mass term with the breaking of hypercubic symmetry in (\ref{eq:8f}) via the spin diagonalization.

\subsection{Central-branch staggered-Wilson fermion}

The symmetry of the four-hopping staggered-Wilson fermion in (\ref{AdamsS1}) is enhanced with the condition $m+r=0$.
The symmetry of $S_{\rm A}$ in (\ref{AdamsS1}) with this condition is
\begin{equation}
\{ C_{0}, \,\,C_{T}'\Xi_{\mu},\,\, C_{T}'I_{s},\,\, R_{\mu\nu} \}\,,
\label{A0Sym}
\end{equation}
where $C_{T}'$ is given as the other special charge conjugation \cite{Steve,Misumi:2012sp,Misumi:2012eh}
\be
C_{T}':\,\,\chi_{x}\,\to\,\bar{\chi}_{x}^{T},\,\,
\bar{\chi}_{x}\,\to\,\chi_{x}^{T},\,\,
U_{x,\mu} \,\to\, U_{x,\mu}^{*}.
\ee
However, the Dirac spectrum has no central branch for this case.

On the other hand, the two-hopping staggered-Wilson fermion in (\ref{HoelS}) with the condition $m+2r=0$
has the central branch in the Dirac spectrum. 
The action with this condition is given by
\begin{align}  
S_{\rm H}^{\rm cb}\,=\, \sum_{xy}\bar{\chi}_{x}(D_{\rm H}^{\rm cb})_{xy}\chi_{y}\,=\,
\sum_{xy}\bar{\chi}_{x}[\eta_{\mu}D_{\mu}
+rM_{\rm H})]_{xy}\chi_{y}\,.
\label{HoelS}
\end{align}
The symmetry \cite{Steve,Misumi:2012sp,Misumi:2012eh} is summarized as
\begin{equation}
\{ C_{T}, \,\,C_{T}',\,\, \Xi_{\mu}', \,\,R_{12},\,\, R_{34},\,\, R_{24}R_{31} \}.
\end{equation}
The extra symmetry is the special charge conjugation $C_{T}'$.
Since the two-flavor central branch exists in the setup,
this enhancement of the symmetry is meaningful.
First of all, the two other mass terms
\be
\bar{\chi}_{x} \chi_{x}\,,\quad\quad
\bar{\chi}_{x} (M_{\rm A})_{xy} \chi_{y} \,,
\ee
are not invariant under the enhanced $C_{T}'$ invariance, thus
their generation by the loop effects is prohibited.
Furthermore, the residual rotational symmetry prohibits unequal renormalization of coefficients of $C_{12}$ and $C_{34}$
in $\bar{\chi}_{x}(M_{\rm H})_{xy} \chi_{y}$.
These facts mean that this two-flavor central-branch fermion is stable in a sense that the additive mass renormalization for each of the two tastes at the central branch is prohibited and the central branch cannot be split by quantum effects.
It is clear difference from the two-flavor central-branch Wilson fermion in Sec.~\ref{sec:2f}, 
but is consistent with the property of the eight-flavor central-branch fermion in Sec.~\ref{sec:Nf}, which is reduced to the central-branch staggered-Wilson fermion by spin diagonalization.
Indeed, the numerical calculation for this case in \cite{deForcrand:2012bm} indicates the absence of additive mass renormalization for the two tastes at the central branch.
We can rephrase this property that the mixed 't Hooft anomaly of the symmetries of the central-branch staggered-Wilson fermion prohibits a trivially gapped phase.

It is notable that the absence of sign problem is also proved in this formulation, where we have $C_{T}'$ instead of the ${\mathbb Z}_{2}$ part of $U(1)_{\overline V}$ in the central-branch Wilson fermion and this $C_{T}'$ leads to the pairing of nonzero eigenvalues in the spectrum of $H\equiv \epsilon_{x} D_{\rm H}^{\rm cb}$. As long as the number of lattice sites is even, the determinant of Dirac operator is positive semi-definite.
When we introduce a mass shift, we can bypass the sign problem easily by quenching the sign of the determinant as proposed in \cite{deForcrand:2012bm}.

Although this two-flavor formulation is free from the necessity of the mass parameter fine-tuning,
we need the one-parameter tuning for restoration of Euclidean Lorentz symmetry.
However, this situation is better than those in the known classes of minimally doubled fermions, where the two- or three-parameter tuning is required \cite{Capitani:2013zta,Capitani:2013iha,Weber:2013tfa,Weber:2017eds,Durr:2020yqa}.
In the practical use of the central-branch staggered-Wilson fermion, one may utilize the knowledge of the anisotropic lattice QCD.



\section{Summary and Discussion}
\label{sec:SD}

In this work we study properties of the several types of central-branch Wilson fermions in four dimensions. We first give a comprehensive review on the original central-branch fermion with introducing several new insights, where we discuss its construction, the prohibition of additive mass renormalization of all six species, spontaneous symmetry breaking in the strong-coupling limit, the absence or the solution of the sign problem for quark determinant, and their practical use.
In particular, we show that, while the sign problem of quark determinant is absent right on the central branch,  the necessity of mass shift in the lattice simulation may revive it. We argue that we can bypass this sign problem just by quenching the sign of the Dirac determinant. We then conclude that the original central-branch Wilson fermion is useful at least in the six or twelve-flavor lattice QCD simulation.

We construct several varieties of the central-branch fermions and study their properties, with special attention to their symmetries.
For instance, we consider the two-flavor version by modifying the Wilson term as ${\sum_{j=1}^{3}C_{j}}+3C_{4}$. Its Dirac spectrum has seven branches and two species live at the central branch. Although the hypercubic symmetry is broken to its cubic subgroup as with the anisotropic lattice formulation, the fermion setup shares all the other symmetries including  C, P, $U(1)_{V}$ and $U(1)_{\overline V}$ with the original central-branch Wilson fermion. 
For this setup, we investigate the additive mass renormalization, the spontaneous symmetry breaking of parity and $U(1)_{\overline V}$ symmetry, the absence of the sign problem, and the parameter tuning for restoration of Euclidean Lorentz symmetry.
In particular, in the lattice perturbation theory we find that the mass difference of the two flavors suffers from additive renormalization due to the breaking of hypercubic symmetry, while the sum of their masses is free from it.
Based on this fact we argue that the existence of full hypercubic symmetry in the original central-branch fermion is essential for the absence of additive mass renormalization for all the six species.
We can rephrase this argument that the mixed 't Hooft anomaly of the symmetries including hypercubic symmetry, $U(1)_{V}$, lattice translation, etc. prohibits a trivially gapped phase.

Other types of central-branch fermions are also discussed, with emphasis on their symmetries. In particular, we investigated the staggered-Wilson fermion as another version of two-flavor central-branch fermions. In the staggered-Wilson fermion, four tastes are split into three branches and two tastes live at the central branch. Thus this fermion formulation without the onsite term is regarded as another version of the two-flavor central-branch fermion. At the central branch, the special type of charge conjugation invariance $C_{T}'$ restores and it prohibits the additive mass renormalization for each of the two tastes. This fermion formulation seems to be a promising two-flavor setup, while the one-parameter tuning for restoration of Euclidean Lorentz symmetry is still required.

The most important messages of this work are summarized as follows:
The original six-flavor central-branch Wilson fermion has enough symmetries to prohibit the additive mass renormalization for all the six species, while it is not the case with the two-flavor version.
Moreover, the central-branch fermions are free from the sign problem right on the central branch, and
it can be bypassed just by the sign quenching even with mass shift.
Thus, the six-flavor central-branch fermion is a promising formulation for six-flavor or twelve-flavor QCD simulations without parameter-tuning.
The central-branch staggered-Wilson fermion also has a stable central branch without any additive mass renormalization, and the one-parameter tuning for Lorentz symmetry in the formulation enables us to study two-flavor QCD efficiently. It can be regarded as a ``minimally doubled fermion" with less symmetry breaking.

In this work, we do not pay much attention to flavor-symmetry breaking among the species living at the central branch. In future works, we have to verify that the whole flavor symmetry restores in lattice QCD with the central-branch fermion in the continuum.
We also plan to perform a full classification of central-branch fermions. It is an interesting question whether it is possible to construct the two-flavor central-branch fermions with full hypercubic symmetry.




\begin{acknowledgements}
T. M. is grateful to Y. Tanizaki for the fruitful discussion on the topics of the paper.
This work of T. M. is supported by the Japan Society for the Promotion of Science (JSPS) 
Grant-in-Aid for Scientific Research (KAKENHI) Grant Numbers 19K03817 and 18H01217.
\end{acknowledgements}


\appendix

\section{5D two-flavor central-branch fermion}
\label{sec:5d}

In five dimensions, we can take a parallel procedure to have two-flavor central-branch fermions.
The deformation from the 5d Wilson is given as
 \begin{equation}
\sum_{\mu=1}^{5} C_{\mu}\,\,\,\to\,\,\, \sum_{j=1}^{4}C_{j}+4C_{5}.
\end{equation}
The free action is just given by
\begin{align}
S_{\rm 5dW2}=\frac{1}{2} \sum _{n}\sum_{\mu=1}^{5} \bar{\psi}_{n}&\gamma _{\mu}(\psi _{n+\hat{\mu}} -\psi _{n-\hat{\mu}} )
\nonumber\\
&-\frac{r}{2}\sum_{n}\bar{\psi}_{n}\left[\sum_{j=1}^{4}\left(\psi_{n+\hat{j}}+\psi_{n-\hat{j}}\right)
+4(\psi_{n+\hat{5}}+\psi_{n-\hat{5}})\right].
\end{align}
The 32 species are split into nine branches with 1, 4, 6, 4, 2, 4, 6, 4 and 1 flavors.
The central branch corresponds to the two zeros of the Dirac operator
$(0,0,0,0,\pi)$ and $(\pi, \pi, \pi, \pi,0)$.
We note that this fermion action explicitly breaks 5d hypercubic symmetry,
while it keeps 4d hypercubic symmetry and other requisite discrete symmetries.

\section{Staggered and staggered-Wilson symmetries}
\label{sec:sym}

In this appendix, we list the staggered discrete symmetries including, staggered charge conjugation, shift, axis reversal and staggered rotation \cite{Golterman:1984cy, Kilcup:1986dg}.
We also discuss their explicit breaking in staggered-Wilson fermions \cite{Steve,Misumi:2012sp,Misumi:2012eh}

The staggered charge-conjugation transformation is given by
\begin{equation}
C_{0}:\,\,\chi_{x}\to\epsilon_{x}\bar{\chi}_{x}^{T},\,\,\,\,
\bar{\chi}_{x}\to-\epsilon_{x}\chi_{x}^{T},\,\,\,\,
U_{x,\nu} \to U_{x,\nu}^{*}\,.
\label{C0}
\end{equation}
The four-hopping flavored-mass term is invariant under this transformation, 
but the two-hopping flavored-mass term is not invariant.

The shift transformation is given by
\begin{align}
\Xi_{\mu}:\,\,\chi_{x} \to \zeta_{\mu}(x)\chi_{x+\hat{\mu}}, \,\,\,\,
\bar{\chi}_{x} \to \zeta_{\mu}(x)\bar{\chi}_{x+\hat{\mu}},\,\,\,\,
U_{x,\nu} \to U_{x+\hat{\mu},\nu}\,,
\label{shift1}
\end{align}
with $\zeta_{1}(x)=(-1)^{x_{2}+x_{3}+x_{4}}$, $\zeta_{2}(x)=(-1)^{x_{3}+x_{4}}$,
$\zeta_{3}(x)=(-1)^{x_{4}}$ and $\zeta_{4}(x)=1$.
This transformation flips the sign of the both types of flavored-mass terms.

The axis reversal transformation is given by
\begin{align}
I_{\mu}:\,\,\chi_{x} \to (-1)^{x_{\mu}}\chi_{x'}, \,\,\,\,
\bar{\chi}_{x} \to (-1)^{x_{\mu}}\bar{\chi}_{x'},\,\,\,\,
U_{x,\nu} \to U_{x',\nu}\,,
\label{axis1}
\end{align}
where $x \to x'$ means $x_{\mu}\to -x_{\mu}$, $x_{\rho}\to x_{\rho}$ with $\rho\not= \mu$. 
In particular, we denote the spatial inversion as $I_{s}=I_{1}I_{2}I_{3}$.
It flips the signs of the both flavored-mass terms.

The staggered rotational transformation is given by
\begin{align}
R_{\rho\sigma}:\,\,\chi_{x} \to S_{R}({\tilde R}^{-1} x)\chi_{{\tilde R}^{-1}x},\,\,\,\,
\bar{\chi_{x}} \to S_{R}({\tilde R}^{-1} x)\bar{\chi}_{{\tilde R}^{-1}x},\,\,\,\,
U_{x,\nu} \to U_{{\tilde R}x,\nu}\,,
\label{rot1}
\end{align}
where $x \to {\tilde R} x$ means $x_{\rho}\to x_{\sigma}$, $x_{\sigma}\to -x_{\rho}$,
$x_{\tau}\to x_{\tau}$, $\tau \not= \rho, \sigma$.
We also define $S_{R}(x)\equiv {1\over{2}}[1\pm\eta_{\rho}(x)\eta_{\sigma}(x)\mp\zeta_{\rho}(x)\zeta_{\sigma}(x)+\eta_{\rho}(x)\eta_{\sigma}(x)\zeta_{\rho}(x)\zeta_{\sigma}(x)]$ with 
$\rho$\hspace{0.3em}\raisebox{0.4ex}{$<$}\hspace{-0.75em}\raisebox{-0.7ex}{$>$}\hspace{0.3em}$\sigma$.
The four-hopping flavored-mass term is invariant under this staggered rotation transformation, 
while the two-hopping type is not.

The physical parity transformation is realized as a combination of $I_{s}$ and $\Xi_{4}$ as
\begin{equation} 
I_{s}\Xi_{4}\chi \,\sim\,   (\gamma_{4}\otimes \id )\psi(-{\bf x},x_{4})\,,
\label{parity}
\end{equation}
where we denote a physical Dirac fermion field as $\psi$.
The two types of staggered-Wilson fermion actions are invariant under this transformation,
thus they have physical parity symmetry P at each of the branches. 
We note the simple combination of $\mu$-shift and $\mu$-axis reversal $I_{\mu} \Xi_{\mu}$ 
is also a symmetry of both staggered-Wilson fermions.

The physical charge conjugation is realized as a combination of staggered charge conjugation,
axis reversal and shift,
\be
C_{0} \Xi_{2}\Xi_{4}I_{2}I_{4}\,\sim\, {\rm C}\,.
\label{eq:C}
\ee
The four-hopping flavored-mass term is invariant under this transformation.
It means that the two flavors at both of the branches have the physical charge conjugation invariance.
Although the two-hopping type breaks $C_{0}$,
it has the other special charge conjugation defined as a combination of $C_{0}$ and the staggered rotation
\begin{equation}
C_{T} : \,\,R_{21}R_{13}^{2}C\,.
\label{RRRC}
\end{equation}
Based on this invariance we can define physical charge conjugation C
for the branches including the central branch. 
Thus, we conclude that fermionic degrees of freedom in both of the staggered-Wilson fermions
have physical charge conjugation invariance.

It is well-known that the diagonal hypercubic transformation $SW_{4, {\rm diag}}$ of euclidian rotation $SO(4)$ and flavor $SU(4)$ is constructed as a combination of the staggered rotation and the axis reversal \cite{Kilcup:1986dg} in staggered fermions
This symmetry is enhanced to Euclidian Lorentz symmetry in the continuum limit. 
The four-hopping staggered-Wilson fermion action is invariant under the staggered rotation and the shifted-axis reversal $\Xi_{\mu}I_{\mu}$, which can form $SW_{4, {\rm diag}}$. 
Thus, the setup is expected to recover Lorentz symmetry in the continuum.
On the other hand, the two-hopping staggered fermion loses the staggered rotation symmetry,
and it results in the necessity of the parameter tuning for the correct continuum limit,

\bibliographystyle{utphys}
\bibliography{./QFT,./refs}

\providecommand{\href}[2]{#2}\begingroup\raggedright\begin{thebibliography}{10}

\bibitem{Wilson:1974sk}
K.~G. Wilson, ``{Confinement of Quarks},''
\href{http://dx.doi.org/10.1103/PhysRevD.10.2445}{{\em Phys. Rev.} {\bfseries
  D10} (1974) 2445--2459}.

\bibitem{Creutz:1980zw}
M.~Creutz, ``{Monte Carlo Study of Quantized SU(2) Gauge Theory},''
\href{http://dx.doi.org/10.1103/PhysRevD.21.2308}{{\em Phys. Rev.} {\bfseries
  D21} (1980) 2308--2315}.

\bibitem{Karsten:1980wd}
L.~H. Karsten and J.~Smit, ``{Lattice Fermions: Species Doubling, Chiral
  Invariance, and the Triangle Anomaly},''
  \href{http://dx.doi.org/10.1016/0550-3213(81)90549-6}{{\em Nucl. Phys.}
  {\bfseries B183} (1981) 103}.
[,495(1980)].

\bibitem{Nielsen:1980rz}
H.~B. Nielsen and M.~Ninomiya, ``{Absence of Neutrinos on a Lattice. 1. Proof
  by Homotopy Theory},''
\href{http://dx.doi.org/10.1016/0550-3213(81)90361-8}{{\em Nucl. Phys.}
  {\bfseries B185} (1981) 20}.

\bibitem{Nielsen:1981xu}
H.~B. Nielsen and M.~Ninomiya, ``{Absence of Neutrinos on a Lattice. 2.
  Intuitive Topological Proof},''
\href{http://dx.doi.org/10.1016/0550-3213(81)90524-1}{{\em Nucl. Phys.}
  {\bfseries B193} (1981) 173--194}.

\bibitem{Nielsen:1981hk}
H.~B. Nielsen and M.~Ninomiya, ``{No Go Theorem for Regularizing Chiral
  Fermions},''
\href{http://dx.doi.org/10.1016/0370-2693(81)91026-1}{{\em Phys. Lett.}
  {\bfseries 105B} (1981) 219--223}.

\bibitem{Wilson:1975id}
K.~G. Wilson, \href{http://dx.doi.org/10.1007/978-1-4613-4208-3_6}{``{Quarks
  and Strings on a Lattice},''} in {\em {New Phenomena in Subnuclear Physics:
  Proceedings, International School of Subnuclear Physics, Erice, Sicily, Jul
  11-Aug 1 1975. Part A}}.
\newblock
1975.
\newblock

\bibitem{Kaplan:1992bt}
D.~B. Kaplan, ``{A Method for simulating chiral fermions on the lattice},''
  \href{http://dx.doi.org/10.1016/0370-2693(92)91112-M}{{\em Phys. Lett.}
  {\bfseries B288} (1992) 342--347},
\href{http://arxiv.org/abs/hep-lat/9206013}{{\ttfamily arXiv:hep-lat/9206013
  [hep-lat]}}.

\bibitem{Shamir:1993zy}
Y.~Shamir, ``{Chiral fermions from lattice boundaries},''
  \href{http://dx.doi.org/10.1016/0550-3213(93)90162-I}{{\em Nucl. Phys.}
  {\bfseries B406} (1993) 90--106},
\href{http://arxiv.org/abs/hep-lat/9303005}{{\ttfamily arXiv:hep-lat/9303005
  [hep-lat]}}.

\bibitem{Furman:1994ky}
V.~Furman and Y.~Shamir, ``{Axial symmetries in lattice QCD with Kaplan
  fermions},'' \href{http://dx.doi.org/10.1016/0550-3213(95)00031-M}{{\em Nucl.
  Phys. B} {\bfseries 439} (1995) 54--78},
  \href{http://arxiv.org/abs/hep-lat/9405004}{{\ttfamily
  arXiv:hep-lat/9405004}}.

\bibitem{Neuberger:1998wv}
H.~Neuberger, ``{More about exactly massless quarks on the lattice},''
  \href{http://dx.doi.org/10.1016/S0370-2693(98)00355-4}{{\em Phys. Lett.}
  {\bfseries B427} (1998) 353--355},
\href{http://arxiv.org/abs/hep-lat/9801031}{{\ttfamily arXiv:hep-lat/9801031
  [hep-lat]}}.

\bibitem{Ginsparg:1981bj}
P.~H. Ginsparg and K.~G. Wilson, ``{A Remnant of Chiral Symmetry on the
  Lattice},''
\href{http://dx.doi.org/10.1103/PhysRevD.25.2649}{{\em Phys. Rev.} {\bfseries
  D25} (1982) 2649}.

\bibitem{Kogut:1974ag}
J.~B. Kogut and L.~Susskind, ``{Hamiltonian Formulation of Wilson's Lattice
  Gauge Theories},''
\href{http://dx.doi.org/10.1103/PhysRevD.11.395}{{\em Phys. Rev.} {\bfseries
  D11} (1975) 395--408}.

\bibitem{Susskind:1976jm}
L.~Susskind, ``{Lattice Fermions},''
\href{http://dx.doi.org/10.1103/PhysRevD.16.3031}{{\em Phys. Rev.} {\bfseries
  D16} (1977) 3031--3039}.

\bibitem{Kawamoto:1981hw}
N.~Kawamoto and J.~Smit, ``{Effective Lagrangian and Dynamical Symmetry
  Breaking in Strongly Coupled Lattice QCD},''.

\bibitem{Sharatchandra:1981si}
H.~S. Sharatchandra, H.~J. Thun, and P.~Weisz, ``{Susskind Fermions on a
  Euclidean Lattice},''
\href{http://dx.doi.org/10.1016/0550-3213(81)90200-5}{{\em Nucl. Phys.}
  {\bfseries B192} (1981) 205--236}.

\bibitem{Golterman:1984cy}
M.~F. Golterman and J.~Smit, ``{Selfenergy and Flavor Interpretation of
  Staggered Fermions},''
  \href{http://dx.doi.org/10.1016/0550-3213(84)90424-3}{{\em Nucl. Phys. B}
  {\bfseries 245} (1984) 61--88}.

\bibitem{Golterman:1985dz}
M.~F. Golterman, ``{STAGGERED MESONS},''
  \href{http://dx.doi.org/10.1016/0550-3213(86)90383-4}{{\em Nucl. Phys. B}
  {\bfseries 273} (1986) 663--676}.

\bibitem{Kilcup:1986dg}
G.~Kilcup and S.~R. Sharpe, ``{A Tool Kit for Staggered Fermions},''
  \href{http://dx.doi.org/10.1016/0550-3213(87)90285-9}{{\em Nucl. Phys. B}
  {\bfseries 283} (1987) 493--550}.

\bibitem{Adams:2009eb}
D.~H. Adams, ``Theoretical foundation for the index theorem on the lattice with
  staggered fermions,''
  \href{http://dx.doi.org/10.1103/PhysRevLett.104.141602}{{\em Phys.Rev.Lett.}
  {\bfseries 104} (2010) 141602},
  \href{http://arxiv.org/abs/0912.2850}{{\ttfamily arXiv:0912.2850 [hep-lat]}}.

\bibitem{Adams:2010gx}
D.~H. Adams, ``Pairs of chiral quarks on the lattice from staggered fermions,''
  \href{http://dx.doi.org/10.1016/j.physletb.2011.04.034}{{\em Phys.Lett.B}
  {\bfseries 699} (2011) 394--397},
  \href{http://arxiv.org/abs/1008.2833}{{\ttfamily arXiv:1008.2833 [hep-lat]}}.

\bibitem{Hoelbling:2010jw}
C.~Hoelbling, ``Single flavor staggered fermions,''
  \href{http://dx.doi.org/10.1016/j.physletb.2010.12.062}{{\em Phys.Lett.B}
  {\bfseries 696} (2011) 422--425},
  \href{http://arxiv.org/abs/1009.5362}{{\ttfamily arXiv:1009.5362 [hep-lat]}}.

\bibitem{deForcrand:2011ak}
P.~de~Forcrand, A.~Kurkela, and M.~Panero, ``{Numerical properties of staggered
  overlap fermions},'' {\em PoS} {\bfseries LATTICE2010} (2010) 080,
  \href{http://arxiv.org/abs/1102.1000}{{\ttfamily arXiv:1102.1000 [hep-lat]}}.

\bibitem{Creutz:2011cd}
M.~Creutz, T.~Kimura, and T.~Misumi, ``{Aoki Phases in the Lattice Gross-Neveu
  Model with Flavored Mass terms},''
  \href{http://dx.doi.org/10.1103/PhysRevD.83.094506}{{\em Phys. Rev.}
  {\bfseries D83} (2011) 094506},
\href{http://arxiv.org/abs/1101.4239}{{\ttfamily arXiv:1101.4239 [hep-lat]}}.

\bibitem{Misumi:2011su}
T.~Misumi, M.~Creutz, T.~Kimura, T.~Z. Nakano, and A.~Ohnishi, ``{Aoki Phases
  in Staggered-Wilson Fermions},''
  \href{http://dx.doi.org/10.22323/1.139.0108}{{\em PoS} {\bfseries
  LATTICE2011} (2011) 108}, \href{http://arxiv.org/abs/1110.1231}{{\ttfamily
  arXiv:1110.1231 [hep-lat]}}.

\bibitem{Follana:2011kh}
E.~Follana, V.~Azcoiti, G.~Di~Carlo, and A.~Vaquero, ``{Spectral Flow and Index
  Theorem for Staggered Fermions},''
  \href{http://dx.doi.org/10.22323/1.139.0100}{{\em PoS} {\bfseries
  LATTICE2011} (2011) 100}, \href{http://arxiv.org/abs/1111.3502}{{\ttfamily
  arXiv:1111.3502 [hep-lat]}}.

\bibitem{deForcrand:2012bm}
P.~de~Forcrand, A.~Kurkela, and M.~Panero, ``Numerical properties of staggered
  quarks with a taste-dependent mass term,''
  \href{http://dx.doi.org/10.1007/JHEP04(2012)142}{{\em JHEP} {\bfseries 04}
  (2012) 142}, \href{http://arxiv.org/abs/1202.1867}{{\ttfamily arXiv:1202.1867
  [hep-lat]}}.

\bibitem{Misumi:2012sp}
T.~Misumi, T.~Z. Nakano, T.~Kimura, and A.~Ohnishi, ``Strong-coupling analysis
  of parity phase structure in staggered-wilson fermions,''
  \href{http://dx.doi.org/10.1103/PhysRevD.86.034501}{{\em Phys.Rev.D}
  {\bfseries 86} (2012) 034501},
  \href{http://arxiv.org/abs/1205.6545}{{\ttfamily arXiv:1205.6545 [hep-lat]}}.

\bibitem{Misumi:2012eh}
T.~Misumi, ``{New fermion discretizations and their applications},''
  \href{http://dx.doi.org/10.22323/1.164.0005}{{\em PoS} {\bfseries
  LATTICE2012} (2012) 005},
\href{http://arxiv.org/abs/1211.6999}{{\ttfamily arXiv:1211.6999 [hep-lat]}}.

\bibitem{Durr:2013gp}
S.~Durr, ``{Taste-split staggered actions: eigenvalues, chiralities and
  Symanzik improvement},''
  \href{http://dx.doi.org/10.1103/PhysRevD.87.114501}{{\em Phys. Rev. D}
  {\bfseries 87} no.~11, (2013) 114501},
  \href{http://arxiv.org/abs/1302.0773}{{\ttfamily arXiv:1302.0773 [hep-lat]}}.

\bibitem{Hoelbling:2016qfv}
C.~Hoelbling and C.~Zielinski, ``{Spectral properties and chiral symmetry
  violations of (staggered) domain wall fermions in the Schwinger model},''
  \href{http://dx.doi.org/10.1103/PhysRevD.94.014501}{{\em Phys. Rev. D}
  {\bfseries 94} no.~1, (2016) 014501},
  \href{http://arxiv.org/abs/1602.08432}{{\ttfamily arXiv:1602.08432
  [hep-lat]}}.

\bibitem{Zielinski:2017pko}
C.~Zielinski, {\em {Theoretical and Computational Aspects of New Lattice
  Fermion Formulations}}.
\newblock PhD thesis, Nanyang Technol. U., 2016.
\newblock \href{http://arxiv.org/abs/1703.06364}{{\ttfamily arXiv:1703.06364
  [hep-lat]}}.

\bibitem{Bietenholz:1999km}
W.~Bietenholz and I.~Hip, ``{The Scaling of exact and approximate
  Ginsparg-Wilson fermions},''
  \href{http://dx.doi.org/10.1016/S0550-3213(99)00477-0}{{\em Nucl. Phys. B}
  {\bfseries 570} (2000) 423--451},
  \href{http://arxiv.org/abs/hep-lat/9902019}{{\ttfamily
  arXiv:hep-lat/9902019}}.

\bibitem{Creutz:2010bm}
M.~Creutz, T.~Kimura, and T.~Misumi, ``{Index Theorem and Overlap Formalism
  with Naive and Minimally Doubled Fermions},''
  \href{http://dx.doi.org/10.1007/JHEP12(2010)041}{{\em JHEP} {\bfseries 12}
  (2010) 041},
\href{http://arxiv.org/abs/1011.0761}{{\ttfamily arXiv:1011.0761 [hep-lat]}}.

\bibitem{Durr:2010ch}
S.~Durr and G.~Koutsou, ``{Brillouin improvement for Wilson fermions},''
  \href{http://dx.doi.org/10.1103/PhysRevD.83.114512}{{\em Phys. Rev. D}
  {\bfseries 83} (2011) 114512},
  \href{http://arxiv.org/abs/1012.3615}{{\ttfamily arXiv:1012.3615 [hep-lat]}}.

\bibitem{Durr:2012dw}
S.~Durr, G.~Koutsou, and T.~Lippert, ``{Meson and Baryon dispersion relations
  with Brillouin fermions},''
  \href{http://dx.doi.org/10.1103/PhysRevD.86.114514}{{\em Phys. Rev. D}
  {\bfseries 86} (2012) 114514},
  \href{http://arxiv.org/abs/1208.6270}{{\ttfamily arXiv:1208.6270 [hep-lat]}}.

\bibitem{Cho:2013yha}
Y.-G. Cho, S.~Hashimoto, J.-I. Noaki, A.~Juttner, and M.~Marinkovic,
  ``{$O(a^2)$-improved actions for heavy quarks and scaling studies on quenched
  lattices},'' \href{http://dx.doi.org/10.22323/1.187.0255}{{\em PoS}
  {\bfseries LATTICE2013} (2014) 255},
  \href{http://arxiv.org/abs/1312.4630}{{\ttfamily arXiv:1312.4630 [hep-lat]}}.

\bibitem{Cho:2015ffa}
Y.-G. Cho, S.~Hashimoto, A.~Juttner, T.~Kaneko, M.~Marinkovic, J.-I. Noaki, and
  J.~T. Tsang, ``{Improved lattice fermion action for heavy quarks},''
  \href{http://dx.doi.org/10.1007/JHEP05(2015)072}{{\em JHEP} {\bfseries 05}
  (2015) 072}, \href{http://arxiv.org/abs/1504.01630}{{\ttfamily
  arXiv:1504.01630 [hep-lat]}}.

\bibitem{Durr:2017wfi}
S.~Durr and G.~Koutsou, ``{On the suitability of the Brillouin action as a
  kernel to the overlap procedure},''
  \href{http://arxiv.org/abs/1701.00726}{{\ttfamily arXiv:1701.00726
  [hep-lat]}}.

\bibitem{Karsten:1981gd}
L.~H. Karsten, ``{Lattice Fermions in Euclidean Space-time},''
  \href{http://dx.doi.org/10.1016/0370-2693(81)90133-7}{{\em Phys. Lett. B}
  {\bfseries 104} (1981) 315--319}.

\bibitem{Wilczek:1987kw}
F.~Wilczek, ``{ON LATTICE FERMIONS},''
  \href{http://dx.doi.org/10.1103/PhysRevLett.59.2397}{{\em Phys. Rev. Lett.}
  {\bfseries 59} (1987) 2397}.

\bibitem{Creutz:2007af}
M.~Creutz, ``{Four-dimensional graphene and chiral fermions},''
  \href{http://dx.doi.org/10.1088/1126-6708/2008/04/017}{{\em JHEP} {\bfseries
  04} (2008) 017}, \href{http://arxiv.org/abs/0712.1201}{{\ttfamily
  arXiv:0712.1201 [hep-lat]}}.

\bibitem{Borici:2007kz}
A.~Borici, ``{Creutz fermions on an orthogonal lattice},''
  \href{http://dx.doi.org/10.1103/PhysRevD.78.074504}{{\em Phys. Rev. D}
  {\bfseries 78} (2008) 074504},
  \href{http://arxiv.org/abs/0712.4401}{{\ttfamily arXiv:0712.4401 [hep-lat]}}.

\bibitem{Bedaque:2008xs}
P.~F. Bedaque, M.~I. Buchoff, B.~C. Tiburzi, and A.~Walker-Loud, ``{Broken
  Symmetries from Minimally Doubled Fermions},''
  \href{http://dx.doi.org/10.1016/j.physletb.2008.03.034}{{\em Phys. Lett. B}
  {\bfseries 662} (2008) 449--455},
  \href{http://arxiv.org/abs/0801.3361}{{\ttfamily arXiv:0801.3361 [hep-lat]}}.

\bibitem{Bedaque:2008jm}
P.~F. Bedaque, M.~I. Buchoff, B.~C. Tiburzi, and A.~Walker-Loud, ``{Search for
  Fermion Actions on Hyperdiamond Lattices},''
  \href{http://dx.doi.org/10.1103/PhysRevD.78.017502}{{\em Phys. Rev. D}
  {\bfseries 78} (2008) 017502},
  \href{http://arxiv.org/abs/0804.1145}{{\ttfamily arXiv:0804.1145 [hep-lat]}}.

\bibitem{Capitani:2009yn}
S.~Capitani, J.~Weber, and H.~Wittig, ``{Minimally doubled fermions at one
  loop},'' \href{http://dx.doi.org/10.1016/j.physletb.2009.09.050}{{\em Phys.
  Lett. B} {\bfseries 681} (2009) 105--112},
  \href{http://arxiv.org/abs/0907.2825}{{\ttfamily arXiv:0907.2825 [hep-lat]}}.

\bibitem{Kimura:2009qe}
T.~Kimura and T.~Misumi, ``{Characters of Lattice Fermions Based on the
  Hyperdiamond Lattice},'' \href{http://dx.doi.org/10.1143/PTP.124.415}{{\em
  Prog. Theor. Phys.} {\bfseries 124} (2010) 415--432},
  \href{http://arxiv.org/abs/0907.1371}{{\ttfamily arXiv:0907.1371 [hep-lat]}}.

\bibitem{Kimura:2009di}
T.~Kimura and T.~Misumi, ``{Lattice Fermions Based on Higher-Dimensional
  Hyperdiamond Lattices},'' \href{http://dx.doi.org/10.1143/PTP.123.63}{{\em
  Prog. Theor. Phys.} {\bfseries 123} (2010) 63--78},
  \href{http://arxiv.org/abs/0907.3774}{{\ttfamily arXiv:0907.3774 [hep-lat]}}.

\bibitem{Creutz:2010cz}
M.~Creutz and T.~Misumi, ``{Classification of Minimally Doubled Fermions},''
  \href{http://dx.doi.org/10.1103/PhysRevD.82.074502}{{\em Phys. Rev. D}
  {\bfseries 82} (2010) 074502},
  \href{http://arxiv.org/abs/1007.3328}{{\ttfamily arXiv:1007.3328 [hep-lat]}}.

\bibitem{Capitani:2010nn}
S.~Capitani, M.~Creutz, J.~Weber, and H.~Wittig, ``{Renormalization of
  minimally doubled fermions},''
  \href{http://dx.doi.org/10.1007/JHEP09(2010)027}{{\em JHEP} {\bfseries 09}
  (2010) 027}, \href{http://arxiv.org/abs/1006.2009}{{\ttfamily arXiv:1006.2009
  [hep-lat]}}.

\bibitem{Tiburzi:2010bm}
B.~C. Tiburzi, ``{Chiral Lattice Fermions, Minimal Doubling, and the Axial
  Anomaly},'' \href{http://dx.doi.org/10.1103/PhysRevD.82.034511}{{\em Phys.
  Rev. D} {\bfseries 82} (2010) 034511},
  \href{http://arxiv.org/abs/1006.0172}{{\ttfamily arXiv:1006.0172 [hep-lat]}}.

\bibitem{Kamata:2011jn}
S.~Kamata and H.~Tanaka, ``{Minimal Doubling Fermion and Hermiticity},''
  \href{http://dx.doi.org/10.1093/ptep/pts093}{{\em PTEP} {\bfseries 2013}
  (2013) 023B05}, \href{http://arxiv.org/abs/1111.4536}{{\ttfamily
  arXiv:1111.4536 [hep-lat]}}.

\bibitem{Misumi:2012uu}
T.~Misumi, ``{Phase structure for lattice fermions with flavored chemical
  potential terms},'' \href{http://dx.doi.org/10.1007/JHEP08(2012)068}{{\em
  JHEP} {\bfseries 08} (2012) 068},
  \href{http://arxiv.org/abs/1206.0969}{{\ttfamily arXiv:1206.0969 [hep-lat]}}.

\bibitem{Misumi:2012ky}
T.~Misumi, T.~Kimura, and A.~Ohnishi, ``{QCD phase diagram with 2-flavor
  lattice fermion formulations},''
  \href{http://dx.doi.org/10.1103/PhysRevD.86.094505}{{\em Phys. Rev. D}
  {\bfseries 86} (2012) 094505},
  \href{http://arxiv.org/abs/1206.1977}{{\ttfamily arXiv:1206.1977 [hep-lat]}}.

\bibitem{Capitani:2013zta}
S.~Capitani, ``{Reducing the number of counterterms with new minimally doubled
  actions},'' \href{http://dx.doi.org/10.1103/PhysRevD.89.014501}{{\em Phys.
  Rev. D} {\bfseries 89} no.~1, (2014) 014501},
  \href{http://arxiv.org/abs/1307.7497}{{\ttfamily arXiv:1307.7497 [hep-lat]}}.

\bibitem{Capitani:2013iha}
S.~Capitani, ``{New chiral lattice actions of the Borici-Creutz type},''
  \href{http://dx.doi.org/10.1103/PhysRevD.89.074508}{{\em Phys. Rev. D}
  {\bfseries 89} no.~7, (2014) 074508},
  \href{http://arxiv.org/abs/1311.5664}{{\ttfamily arXiv:1311.5664 [hep-lat]}}.

\bibitem{Misumi:2013maa}
T.~Misumi, ``{Fermion Actions extracted from Lattice Super Yang-Mills
  Theories},'' \href{http://dx.doi.org/10.1007/JHEP12(2013)063}{{\em JHEP}
  {\bfseries 12} (2013) 063}, \href{http://arxiv.org/abs/1311.4365}{{\ttfamily
  arXiv:1311.4365 [hep-lat]}}.

\bibitem{Weber:2013tfa}
J.~H. Weber, S.~Capitani, and H.~Wittig, ``{Numerical studies of Minimally
  Doubled Fermions},'' \href{http://dx.doi.org/10.22323/1.187.0122}{{\em PoS}
  {\bfseries LATTICE2013} (2014) 122},
  \href{http://arxiv.org/abs/1312.0488}{{\ttfamily arXiv:1312.0488 [hep-lat]}}.

\bibitem{Weber:2017eds}
J.~H. Weber, {\em {Properties of minimally doubled fermions}}.
\newblock PhD thesis, Mainz U., 2015.
\newblock \href{http://arxiv.org/abs/1706.07104}{{\ttfamily arXiv:1706.07104
  [hep-lat]}}.

\bibitem{Durr:2020yqa}
S.~Durr and J.~H. Weber, ``{Dispersion relation and spectral range of
  Karsten-Wilczek and Borici-Creutz fermions},''
  \href{http://arxiv.org/abs/2003.10803}{{\ttfamily arXiv:2003.10803
  [hep-lat]}}.

\bibitem{Kimura:2011ik}
T.~Kimura, S.~Komatsu, T.~Misumi, T.~Noumi, S.~Torii, and S.~Aoki,
  ``{Revisiting symmetries of lattice fermions via spin-flavor
  representation},'' \href{http://dx.doi.org/10.1007/JHEP01(2012)048}{{\em
  JHEP} {\bfseries 01} (2012) 048},
\href{http://arxiv.org/abs/1111.0402}{{\ttfamily arXiv:1111.0402 [hep-lat]}}.

\bibitem{Chowdhury:2013ux}
A.~Chowdhury, A.~Harindranath, J.~Maiti, and S.~Mondal, ``Many avatars of the
  wilson fermion: A perturbative analysis,''
  \href{http://dx.doi.org/10.1007/JHEP02(2013)037}{{\em JHEP} {\bfseries 02}
  (2013) 037}, \href{http://arxiv.org/abs/1301.0675}{{\ttfamily arXiv:1301.0675
  [hep-lat]}}.

\bibitem{Misumi:2019jrt}
T.~Misumi and Y.~Tanizaki, ``{Lattice gauge theory for Haldane conjecture and
  central-branch Wilson fermion},''
  \href{http://dx.doi.org/10.1093/ptep/ptaa003}{{\em PTEP} {\bfseries 2020}
  no.~3, (2020) 033B03}, \href{http://arxiv.org/abs/1910.09604}{{\ttfamily
  arXiv:1910.09604 [hep-lat]}}.

\bibitem{tHooft:1979rat}
G.~'t~Hooft,
  \href{http://dx.doi.org/10.1007/978-1-4684-7571-5_9}{``{Naturalness, chiral
  symmetry, and spontaneous chiral symmetry breaking},''} in {\em {Recent
  Developments in Gauge Theories. Proceedings, Nato Advanced Study Institute,
  Cargese, France, August 26 - September 8, 1979}}, vol.~59, pp.~135--157.
\newblock
1980.
\newblock

\bibitem{Frishman:1980dq}
Y.~Frishman, A.~Schwimmer, T.~Banks, and S.~Yankielowicz, ``{The Axial Anomaly
  and the Bound State Spectrum in Confining Theories},''
\href{http://dx.doi.org/10.1016/0550-3213(81)90268-6}{{\em Nucl. Phys.}
  {\bfseries B177} (1981) 157--171}.

\bibitem{Wen:2013oza}
X.-G. Wen, ``{Classifying gauge anomalies through symmetry-protected trivial
  orders and classifying gravitational anomalies through topological orders},''
  \href{http://dx.doi.org/10.1103/PhysRevD.88.045013}{{\em Phys. Rev.}
  {\bfseries D88} no.~4, (2013) 045013},
\href{http://arxiv.org/abs/1303.1803}{{\ttfamily arXiv:1303.1803 [hep-th]}}.

\bibitem{Kapustin:2014lwa}
A.~Kapustin and R.~Thorngren, ``{Anomalies of discrete symmetries in three
  dimensions and group cohomology},''
  \href{http://dx.doi.org/10.1103/PhysRevLett.112.231602}{{\em Phys. Rev.
  Lett.} {\bfseries 112} no.~23, (2014) 231602},
\href{http://arxiv.org/abs/1403.0617}{{\ttfamily arXiv:1403.0617 [hep-th]}}.

\bibitem{Cho:2014jfa}
G.~Y. Cho, J.~C.~Y. Teo, and S.~Ryu, ``{Conflicting Symmetries in Topologically
  Ordered Surface States of Three-dimensional Bosonic Symmetry Protected
  Topological Phases},''
  \href{http://dx.doi.org/10.1103/PhysRevB.89.235103}{{\em Phys. Rev.}
  {\bfseries B89} no.~23, (2014) 235103},
\href{http://arxiv.org/abs/1403.2018}{{\ttfamily arXiv:1403.2018
  [cond-mat.str-el]}}.

\bibitem{Wang:2014pma}
J.~C. Wang, Z.-C. Gu, and X.-G. Wen, ``{Field theory representation of
  gauge-gravity symmetry-protected topological invariants, group cohomology and
  beyond},'' \href{http://dx.doi.org/10.1103/PhysRevLett.114.031601}{{\em Phys.
  Rev. Lett.} {\bfseries 114} no.~3, (2015) 031601},
\href{http://arxiv.org/abs/1405.7689}{{\ttfamily arXiv:1405.7689
  [cond-mat.str-el]}}.

\bibitem{Witten:2016cio}
E.~Witten, ``{The "Parity" Anomaly On An Unorientable Manifold},''
  \href{http://dx.doi.org/10.1103/PhysRevB.94.195150}{{\em Phys. Rev.}
  {\bfseries B94} no.~19, (2016) 195150},
\href{http://arxiv.org/abs/1605.02391}{{\ttfamily arXiv:1605.02391 [hep-th]}}.

\bibitem{Tachikawa:2016cha}
Y.~Tachikawa and K.~Yonekura, ``{On time-reversal anomaly of 2+1d topological
  phases},'' \href{http://dx.doi.org/10.1093/ptep/ptx010}{{\em PTEP} {\bfseries
  2017} no.~3, (2017) 033B04},
\href{http://arxiv.org/abs/1610.07010}{{\ttfamily arXiv:1610.07010 [hep-th]}}.

\bibitem{Gaiotto:2017yup}
D.~Gaiotto, A.~Kapustin, Z.~Komargodski, and N.~Seiberg, ``{Theta, Time
  Reversal, and Temperature},''
  \href{http://dx.doi.org/10.1007/JHEP05(2017)091}{{\em JHEP} {\bfseries 05}
  (2017) 091},
\href{http://arxiv.org/abs/1703.00501}{{\ttfamily arXiv:1703.00501 [hep-th]}}.

\bibitem{Tanizaki:2017bam}
Y.~Tanizaki and Y.~Kikuchi, ``{Vacuum structure of bifundamental gauge theories
  at finite topological angles},''
  \href{http://dx.doi.org/10.1007/JHEP06(2017)102}{{\em JHEP} {\bfseries 06}
  (2017) 102},
\href{http://arxiv.org/abs/1705.01949}{{\ttfamily arXiv:1705.01949 [hep-th]}}.

\bibitem{Shimizu:2017asf}
H.~Shimizu and K.~Yonekura, ``{Anomaly constraints on deconfinement and chiral
  phase transition},'' \href{http://dx.doi.org/10.1103/PhysRevD.97.105011}{{\em
  Phys. Rev.} {\bfseries D97} no.~10, (2018) 105011},
\href{http://arxiv.org/abs/1706.06104}{{\ttfamily arXiv:1706.06104 [hep-th]}}.

\bibitem{Tanizaki:2017qhf}
Y.~Tanizaki, T.~Misumi, and N.~Sakai, ``{Circle compactification and 't Hooft
  anomaly},'' \href{http://dx.doi.org/10.1007/JHEP12(2017)056}{{\em JHEP}
  {\bfseries 12} (2017) 056},
\href{http://arxiv.org/abs/1710.08923}{{\ttfamily arXiv:1710.08923 [hep-th]}}.

\bibitem{Sulejmanpasic:2018upi}
T.~Sulejmanpasic and Y.~Tanizaki, ``{C-P-T anomaly matching in bosonic quantum
  field theory and spin chains},''
  \href{http://dx.doi.org/10.1103/PhysRevB.97.144201}{{\em Phys. Rev.}
  {\bfseries B97} (2018) 144201},
\href{http://arxiv.org/abs/1802.02153}{{\ttfamily arXiv:1802.02153 [hep-th]}}.

\bibitem{Tanizaki:2018xto}
Y.~Tanizaki and T.~Sulejmanpasic, ``{Anomaly and global inconsistency matching:
  $\theta$-angles, $SU(3)/U(1)^2$ nonlinear sigma model, $SU(3)$ chains and its
  generalizations},'' \href{http://dx.doi.org/10.1103/PhysRevB.98.115126}{{\em
  Phys. Rev.} {\bfseries B98} no.~11, (2018) 115126},
\href{http://arxiv.org/abs/1805.11423}{{\ttfamily arXiv:1805.11423
  [cond-mat.str-el]}}.

\bibitem{Svetitsky:1980pa}
B.~Svetitsky, S.~Drell, H.~R. Quinn, and M.~Weinstein, ``{Dynamical Breaking of
  Chiral Symmetry in Lattice Gauge Theories},''
  \href{http://dx.doi.org/10.1103/PhysRevD.22.490}{{\em Phys. Rev. D}
  {\bfseries 22} (1980) 490}.

\bibitem{Blairon:1980pk}
J.~Blairon, R.~Brout, F.~Englert, and J.~Greensite, ``{Chiral Symmetry Breaking
  in the Action Formulation of Lattice Gauge Theory},''
  \href{http://dx.doi.org/10.1016/0550-3213(81)90061-4}{{\em Nucl. Phys. B}
  {\bfseries 180} (1981) 439--457}.

\bibitem{Aoki:1983qi}
S.~Aoki, ``{New Phase Structure for Lattice QCD with Wilson Fermions},''
\href{http://dx.doi.org/10.1103/PhysRevD.30.2653}{{\em Phys. Rev.} {\bfseries
  D30} (1984) 2653}.

\bibitem{Aoki:1986kt}
S.~Aoki, ``{The U(1) Problem on a Lattice. 2. Strong Coupling Expansion},''
  \href{http://dx.doi.org/10.1103/PhysRevD.34.3170}{{\em Phys. Rev. D}
  {\bfseries 34} (1986) 3170}.

\bibitem{Aoki:1986xr}
S.~Aoki, ``{A Solution to the U(1) Problem on a Lattice},''
\href{http://dx.doi.org/10.1103/PhysRevLett.57.3136}{{\em Phys. Rev. Lett.}
  {\bfseries 57} (1986) 3136}.

\bibitem{Aoki:1987us}
S.~Aoki, ``{U(1) Problem and Lattice QCD},''
\href{http://dx.doi.org/10.1016/0550-3213(89)90113-2}{{\em Nucl. Phys.}
  {\bfseries B314} (1989) 79--111}.

\bibitem{Sharpe:1998xm}
S.~R. Sharpe and J.~Singleton, Robert~L., ``{Spontaneous flavor and parity
  breaking with Wilson fermions},''
  \href{http://dx.doi.org/10.1103/PhysRevD.58.074501}{{\em Phys. Rev. D}
  {\bfseries 58} (1998) 074501},
  \href{http://arxiv.org/abs/hep-lat/9804028}{{\ttfamily
  arXiv:hep-lat/9804028}}.

\bibitem{Creutz:1996bg}
M.~Creutz, ``{Wilson fermions at finite temperature},'' in {\em {RHIC Summer
  Study 96: Brookhaven Theory Workshop on Relativistic Heavy Ions}}.
\newblock 7, 1996.
\newblock \href{http://arxiv.org/abs/hep-lat/9608024}{{\ttfamily
  arXiv:hep-lat/9608024}}.

\bibitem{Sint:2007ug}
S.~Sint, \href{http://dx.doi.org/10.1142/9789812790927\_0004}{``{Lattice QCD
  with a chiral twist},''} in {\em {Workshop on Perspectives in Lattice QCD}}.
\newblock 2, 2007.
\newblock \href{http://arxiv.org/abs/hep-lat/0702008}{{\ttfamily
  arXiv:hep-lat/0702008}}.

\bibitem{Shindler:2007vp}
A.~Shindler, ``{Twisted mass lattice QCD},''
  \href{http://dx.doi.org/10.1016/j.physrep.2008.03.001}{{\em Phys. Rept.}
  {\bfseries 461} (2008) 37--110},
  \href{http://arxiv.org/abs/0707.4093}{{\ttfamily arXiv:0707.4093 [hep-lat]}}.

\bibitem{Morrin:2006tf}
R.~Morrin, A.~O. Cais, M.~Peardon, S.~M. Ryan, and J.-I. Skullerud,
  ``{Dynamical QCD simulations on anisotropic lattices},''
  \href{http://dx.doi.org/10.1103/PhysRevD.74.014505}{{\em Phys. Rev. D}
  {\bfseries 74} (2006) 014505},
  \href{http://arxiv.org/abs/hep-lat/0604021}{{\ttfamily
  arXiv:hep-lat/0604021}}.

\bibitem{Lin:2008pr}
{\bfseries Hadron Spectrum} Collaboration, H.-W. Lin {\em et~al.}, ``{First
  results from 2+1 dynamical quark flavors on an anisotropic lattice:
  Light-hadron spectroscopy and setting the strange-quark mass},''
  \href{http://dx.doi.org/10.1103/PhysRevD.79.034502}{{\em Phys. Rev. D}
  {\bfseries 79} (2009) 034502},
  \href{http://arxiv.org/abs/0810.3588}{{\ttfamily arXiv:0810.3588 [hep-lat]}}.

\bibitem{M1}
T.~Misumi. {PhD thesis, Kyoto University}, 2012.
\newblock http://hdl.handle.net/2433/157773.

\bibitem{Steve}
S.~Sharpe. {Talk at YIPQS-HPCI workshop ``New-Type of Fermions on the
  Lattice"}, 2012.
\newblock
  http://www2.yukawa.kyoto-u.ac.jp/ws/2011/newtype/Talk-slides/sharpe-kyoto12-1.pd.

\end{thebibliography}\endgroup

\end{document}